\documentclass[a4paper,fleqn,usenatbib,useAMS]{mnras}

\usepackage{graphicx}	
\usepackage{amsmath}	
\usepackage{amssymb}	
\usepackage{multicol}        
\usepackage{bm}		
\usepackage{pdflscape}	

\usepackage[T1]{fontenc}
\usepackage{ae,aecompl}

\usepackage{newtxtext,newtxmath}

\title[C-rich dust in A30 \& A78]{Carbon dust in the evolved born-again planetary nebulae A\,30 and A\,78}

\author[J.A.\,Toal\'{a} et al.]{J.A.\,Toal\'{a}$^{1}$\thanks{E-mail:\,j.toala@irya.unam.mx}, P.\,Jim\'{e}nez-Hern\'{a}ndez$^{1}$, J.B.\,Rodr\'{i}guez-Gonz\'{a}lez$^{1}$,  S.\,Estrada-Dorado$^{1}$, M.A.\,Guerrero$^{2}$
\newauthor V.M.A.\,G\'{o}mez-Gonz\'{a}lez$^{1}$,  G.\,Ramos-Larios$^{3,4}$, D.A.\,Garc\'{i}a-Hern\'{a}ndez$^{5,6}$ and H.\,Todt$^{7}$\\
$^{1}$Instituto de Radioastronom\'{i}a y Astrof\'{i}sica, UNAM Campus Morelia, Apartado postal 3-72, 58090 Morelia, Mich., Mexico\\
$^{2}$Instituto de Astrof\'{i}sica de Andaluc\'{i}a, IAA-CSIC, Glorieta de la Astronom\'{i}a S/N, Granada 18008, Spain\\
$^{3}$CUCEI, Universidad de Guadalajara, Blvd. Marcelino Garc\'\i a Barrag\'an 1421, 44430, Guadalajara, Jalisco, Mexico \\
$^{4}$Instituto de Astronom\'\i a y Meteorolog\'\i a, Dpto.\ de F\'\i sica,CUCEI, Av.\ Vallarta 2602, 44130, Guadalajara, Jalisco, Mexico\\
$^{5}$Instituto de Astrof\'{i}sica de Canarias, C/ Via Lactea s/n, E-38200 La Laguna, Spain\\
$^{6}$Universidad de La Laguna (ULL), Departamento de Astrofísica, E-38206 La Laguna, Tenerife, Spain\\
$^{7}$Institute for Physics and Astronomy, Universit\"{a}t Potsdam, D-14476 Potsdam, Germany
}

\pubyear{2020}

\begin{document}
\label{firstpage}
\pagerange{\pageref{firstpage}--\pageref{lastpage}}
\maketitle

\begin{abstract}
We present an infrared (IR) characterization of the born-again planetary nebulae (PNe) A\,30
and A\,78 using IR images and spectra. We demonstrate that the carbon-rich dust in A\,30 and A\,78 is spatially coincident with the 
H-poor ejecta and coexists with hot X-ray-emitting gas up to distances of 50$''$ from the central stars (CSPNs). Dust forms immediately after the born-again event and survives for 1000~yr in the harsh environment around the CSPN as it is destroyed and pushed away by radiation pressure and dragged by hydrodynamical effects. {\it Spitzer} IRS spectral maps showed that the broad spectral features at 6.4 and 8.0~$\mu$m, attributed to amorphous carbon formed in H-deficient environments, are associated with the disrupted disk around their CSPN, providing an optimal environment for charge exchange reactions with the stellar wind that produces the soft X-ray emission of these sources. Nebular and dust properties are modeled for A\,30 with {\sc cloudy} taking into account different 
carbonaceous dust species. Our models predict dust temperatures in the 40--230~K range, five times lower than predicted by previous works. Gas and dust masses for the born-again ejecta in A\,30 are estimated to be $M_\mathrm{gas}=(4.41^{+0.55}_{-0.14})\times10^{-3}$~M$_\odot$ 
and $M_\mathrm{dust}=(3.20^{+3.21}_{-2.06})\times10^{-3}$~M$_\odot$, which can be used to estimate a total ejected mass and mass-loss rate for the born-again event of $(7.61^{+3.76}_{-2.20})\times10^{-3}$~M$_{\odot}$ and $\dot{M}=[5-60]\times10^{-5}$~M$_{\odot}$~yr$^{-1}$, respectively. Taking into account the carbon trapped into dust grains, we estimate that the C/O mass ratio of the H-poor ejecta of A\,30 is larger than 1, which favors the very late thermal pulse model over the alternate hypothesis of a nova-like event.

\end{abstract}





\section{INTRODUCTION}
\label{sec:intro}

{\it Born-again} planetary nebulae (PNe) represent a very specific
case in the evolution of low- and intermediate-mass stars
($M_\mathrm{i} \lesssim 1-8$~M$_{\odot}$).
By the late 1970s and early 1980s, theoretical works proposed that the H-rich outer layer of a central star of a PN (CSPN) could build a shell of helium with the critical mass to ignite into carbon and oxygen \citep[see][]{Schonberner1979,Iben1983}, producing what is known today as a {\it very late   thermal pulse} \citep[VLTP; e.g.,][]{Herwig1999}. 
According to this process, an explosive event would eject newly processed,  metal-rich material inside the old hydrogen-rich PN while leaving the central star intact. 
Models
such as those presented by \citet{MM2006} predict that for a short
period of time ($\sim$20--100~yr), the star returns to the asymptotic
giant branch (AGB) region of the HR diagram 
to finally evolve into a second white dwarf phase developing a new 
fast stellar wind and increasing its ionizing photon flux. 
The result is what it is commonly know as a born-again PN.
Although the number of reported born-again PNe is small 
\citep[see][for the latest discoveries]{Guerrero2018,Gv2020}, they give the
opportunity to study complex effects which take place in human-time scales.

The most studied object of this class is the one known as Sakurai's Object (a.k.a.\ V\,4334\,Sgr), which experienced an outburst in 1996 \citep[][]{Nakano1996}. 
Sakurai's Object is surrounded by a hydrogen-rich outer nebula with an angular diameter of $\sim$40$''$ \citep[e.g.,][]{Duerbeck1996,Pollacco1999}, but early studies showed a considerable change in composition consistent with the ingestion of the hydrogen shell and the generation of $s$-process elements \citep[][]{Asplund1997}. 
Within $\sim$20~yr, the CSPN of Sakurai's Object and its ejecta have exhibited dramatic changes \citep[e.g.,][]{Evans2020}. 
The CSPN varied its flux from $K\sim$11.4~mag in 1996 February to  $K\sim25$~mag by 1999, becoming obscured by dust and no longer detectable in optical wavelengths \citep[][]{Duerbeck2000}. Very Large Array (VLA) observations pointed at the reionization of the central core of Sakurai's Object by 2004 \citep[see][and references therein]{vanHoof2007}, yet the dust shell was still cooling by 2005 at temperatures $\sim$200~K according to {\it Spitzer} IRS observations \citep{Evans2006}.  
The ejecta has kept expanding since then, decreasing its optical thickness as revealed by the increase in magnitude up to $K_\mathrm{s}=13.8$~mag by 2014 June \citep{Hinkle2015}, until it has been finally detected in near-IR observations a bipolar ejecta being blown by the current increase of the stellar wind from the CSPN of Sakurai's Object. 
Sub-millimeter observations report that the hydrogen-poor ejecta is currently moving at velocities $\gtrsim 100$~km~s$^{-1}$ \citep[see, e.g.,][]{Tafoya2017}.

The detailed morphology and angular expansion of the more evolved born-again PNe A\,30, A\,58 and A\,78 have been investigated using {\it Hubble Space Telescope (HST)} observations \citep{Clayton2013,Fang2014}. 
Similarly to Sakurai's Object, these three objects exhibit a nearly spherical outer hydrogen-rich PNe, but their VLTP ejecta are rather asymmetric.
A\,58, which experienced a VLTP around 100~yr ago, exhibits the clear presence of a hydrogen-deficient disk which is being disrupted by the current fast wind \citep[$v_{\infty}\approx2500$ km~s$^{-1}$;][]{Clayton2006} and the increased UV flux from its CSPN \citep[][]{Hinkle2008}. 
A\,30 and A\,78 are the oldest objects of this class with a time-lapse since the VLTP event  $\lesssim1000$~yr \citep{Fang2014}. 
Hydrogen-poor structures were discovered around their CSPNe (see Fig.~\ref{fig:fig1} left panels) by \citet{Jacoby1979} and subsequent high resolution {\it HST} observations \citep[][]{Borkowski1993,Borkowski1995} revealed that these seem to be arranged in disrupted disks and bipolar outflows orthogonal to them \citep[see Fig.~\ref{fig:A30} and \ref{fig:A78} upper left panels; e.g,][and references therein]{Meaburn1998,Fang2014}. 
These morphologies have been used to argue that their CSPNe might be binary systems. 
Indeed, \citet{Jacoby2020} has recently reported periodic variations in the light curve of the CSPN of A\,30, which they attribute to  the presence of a binary system.

\subsection{Dust formation in born-again PNe}

As a consequence of the VLTP, the CSPN moves back to the AGB region of the HR diagram for a short period of time. The significant reduction of the effective temperature ($T_\mathrm{eff}$) allows the formation of dust from the H-poor, C-enhanced material that will be injected into the old PN. \citet{Evans2020} showed the dramatic evolution of the dust and molecules in Sakurai's Object in a time period of only 20~yr, when the dust in its born-again ejecta cooled down from more than 1000~K to $\sim$180~K and its mass increased from 10$^{-10}$~M$_{\odot}$ to 10$^{-5}$~M$_{\odot}$. On the other hand, the 100~yr old ejecta in A\,58 has been reported to harbor dust with a range of temperatures, $\lesssim$10$^{-5}$ M$_\odot$ of warm dust with temperatures in the 200--800~K range and 2$\times$10$^{-3}$ M$_\odot$ of cold dust with a temperature of 75~K \citep{Clayton2013}. The latter is in the range of the dust mass of $8\times10^{-3}$~M$_\odot$ estimated by \citet{Koller2001}.

The exceptional dust enrichment of A\,30 and A\,78 compared to other 
PNe was first noted in the analysis of IR photometric observations 
in the 2.2--22~$\mu$m wavelength range of 113 PNe presented by 
\citet{Cohen1974}. The photometry of A\,30 suggested that most 
of the IR emission did not arise from the CSPN, but from an 
extended $\sim25-30''$ in diameter region within the main nebula 
\citep[see also][]{Moseley1980}. This conclusion was supported by 
subsequent work presented by \citet{Cohen1977} who studied optical, 
near- and mid-IR observations of A\,30 and A\,78 and concluded that 
thermal emission from dust should be extended with dust present up 
to distances as far as $\gtrsim10-30''$ from their CSPNe. 
The first ever direct IR image of any born-again PN presented by 
\citet{Dinerstein1984} corroborated the suggestion of previous 
works that the dust distribution is clumpy. Moreover, their $K$ 
and $N$ near-IR images of the central region of A\,30 suggested 
that dust was mainly distributed along a disrupted disk, although 
the spatial resolution of the observations was not optimal.

\begin{figure*}
\begin{center}
  \includegraphics[angle=0,width=\linewidth]{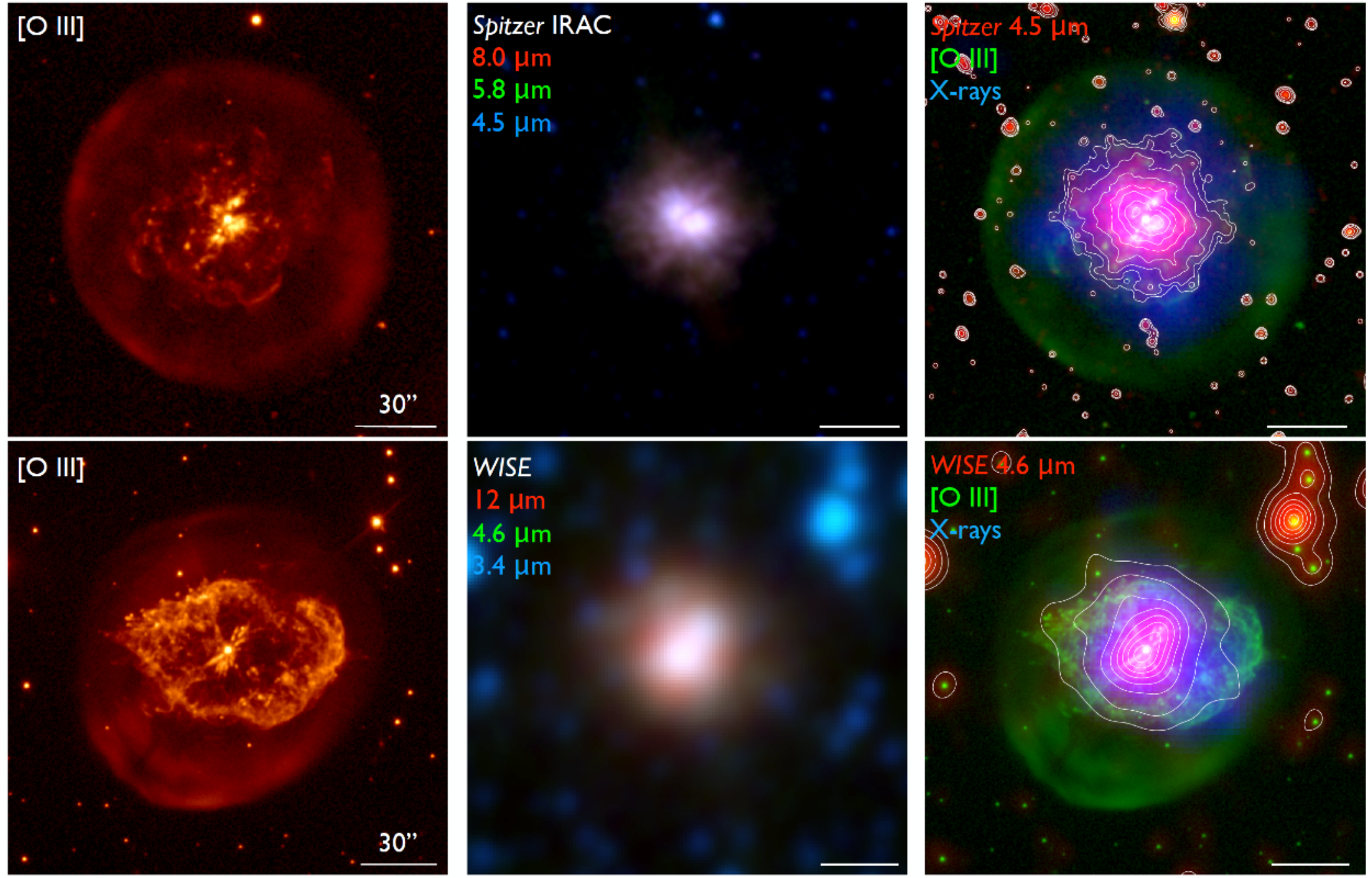}
  \label{fig:largescale}
\caption{
Comparison between the IR, optical and X-ray emission of A\,30 (top panels) and A\,78 (bottom panels). 
The left panels show optical [O~{\sc iii}] images obtained at the Kitt Peak National Observatory Mayall telescope for A\,30 and the Nordic Optical Telescope (NOT) for A\,78, the middle panels show colour-composite IR images, and the right panels show colour-composite images combining IR (red), optical [O\,{\sc iii}] (green) and {\it XMM-Newton} soft X-ray  (0.3--0.5~keV; blue) images. 
To facilitate the comparison we show the IR contours. 
In all images North is up, East to the left.
}
\label{fig:fig1}
\end{center}
\end{figure*}

The first high-resolution near-IR $K$-band image of A\,30 resolved in 
great detail the hot dust of its inner region, which is spatially 
associated with the disrupted disk around the CSPN \citep{Borkowski1994}. 
These authors used all available IR photometry covering the wavelength 
range from 1.2 to 25~$\mu$m \citep[][]{Pottasch1984,Leene1988} and 
produced models of thermally-emitting carbon rich dust.  
The smallest size of the dust grains in these models had to be 
0.0007~$\mu$m in order to fit the IR spectral energy distribution 
(SED) around 1--5~$\mu$m, with a complete size distribution of 
0.0007--0.25~$\mu$m. Furthermore, the presence of graphite and 
polycyclic aromatic hydrocarbons (PAHs) in the dust of A\,30 was rejected 
and calculations performed using only amorphous carbon grains. 
The final model presented by \citet{Borkowski1994} includes 
contribution from two concentric shells, one accounting for the 
disrupted disk and jets and another for the extended clumpy 
distribution around A\,30 (see Fig.~\ref{fig:fig1} top left panel). 
This model predicts a dust mass of $3.6\times10^{-5}$~M$_\odot$ 
for the inner shell and about 60 times more for the outer shell, 
$2.2\times10^{-3}$~M$_\odot$.

Recently, \citet{Muthu2020} presented a statistical analysis of the IR properties of the PNe with Wolf-Rayet-type CSPN compiled by \citet{Weidmann2011} using photometric measurements from the {\it Two Micron All Sky Survey (2MASS)}, {\it Wide-field Infrared Survey (WISE}), {\it Infrared Astronomical Satellite (IRAS}), and {\it Akari}. This study computed the dust colour temperatures, dust masses, and dust-to-gas mass ratios, among other quantities of these PNe. 
They report a dust mass of $10^{-4}$~M$_\odot$ for A\,30, with a dust-to-gas ratio $>$30.  
No values were reported for A\,78.

In this paper we present the analysis of IR images and spectra of A\,30 and A\,78. 
The images reveal in great detail the spatial distribution of multiple dust components in these PNe,  
whereas the spectra inform about their mid-IR spectral properties. 
A model of the gas and dust tailored specifically to A\,30 has been used 
to estimate its total mass ejected during the VLTP event. 
This paper is organized as follows. 
The observations are described in Section~2. 
The distribution of dust and its spectral properties are presented in Sections~3 and 4, respectively. 
Our dust model and its predictions are introduced in Section~5. 
We discuss our results in Section~6 and present our summary and conclusions in Section~7.

\section{Observations}

\begin{figure*}
\begin{center}
\includegraphics[angle=0,width=\linewidth]{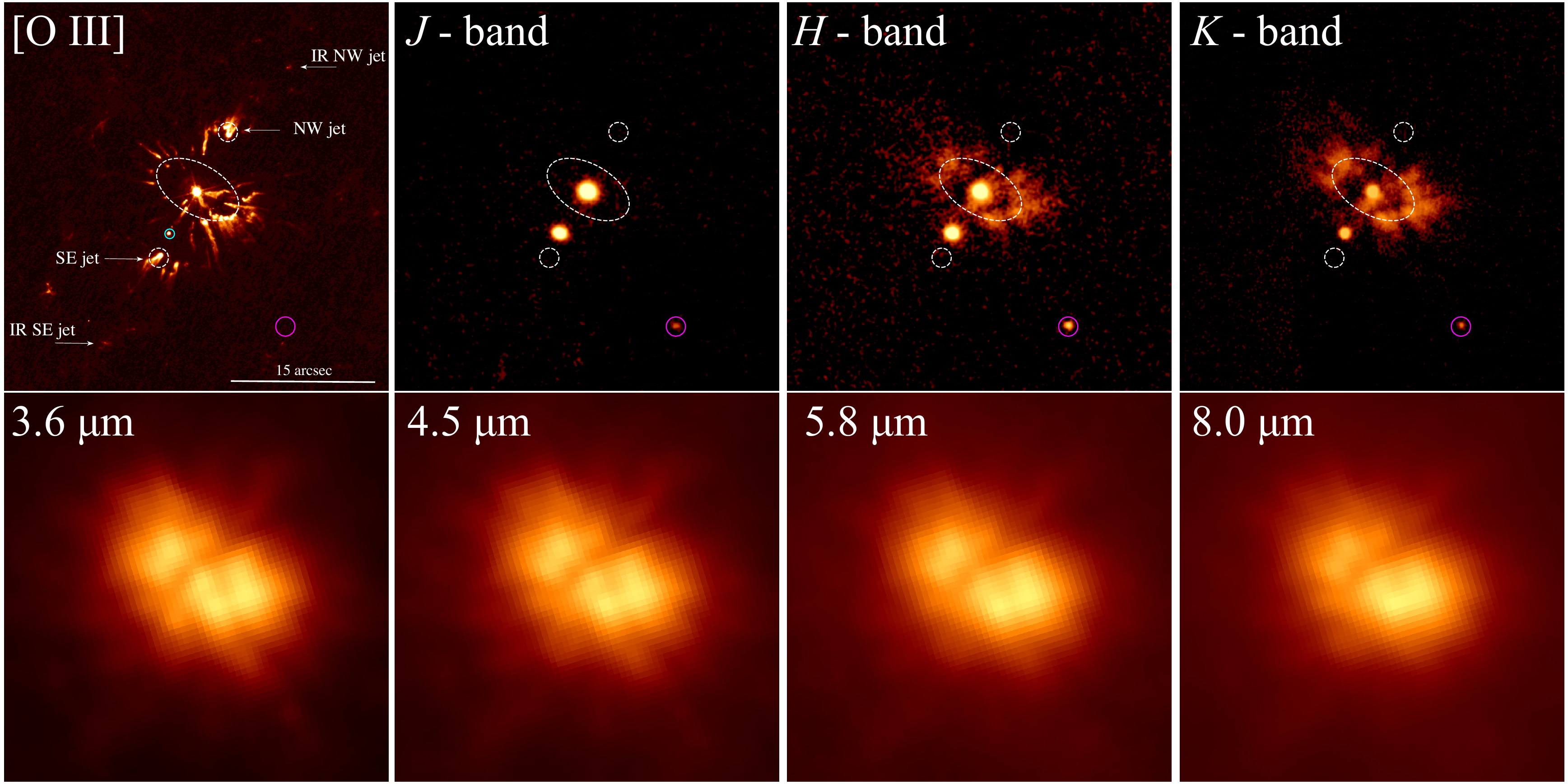}
\label{fig:A30}
\caption{
The inner region of A\,30 as seen in different wavelengths.  
The top panels show the {\it HST} [O\,{\sc iii}] narrow-band (leftmost panel) and NOTCam $J$, $H$ and $K$ images, whereas the bottom panels show \emph{Spitzer} IRAC 3.6, 4.5, 5.8 and 8.0~$\mu$m images. 
The disrupted disk extension is shown with a dashed-line ellipse while the positions of the jet features are marked by dashed-line circles in the top panels. 
A background star is shown with cyan circle. 
The position of a clump that appears in the NOTCam images is shown with a magenta circle. 
All panels correspond to the same FoV. 
North is up, East to the left.
}
\label{fig:A30}
\end{center}
\end{figure*}

IR observations of A\,30 and A\,78 were obtained from different
telescopes. Archival {\it Spitzer} IRS and IRAC observations as well
as {\it WISE} images were obtained from the NASA/IPAC Infrared Science
Archive\footnote{\url{https://irsa.ipac.caltech.edu/frontpage/}}. A\,30
was observed by {\it Spitzer} on 2007 November 25 and December 11
in the staring mode. The IRAC and IRS observations correspond to 
Program ID.\,40115 (PI:G.\,Fazio; AORKey:\,21967616, 21967872 and
21968128). The {\it Spitzer} IRS observations of A\,78 used here
correspond to Program ID.\,3362 (PI:\,A.\,Evans;
AORKey:\,10839808) and were obtained on 2004 November 6, 14 and
17 in mapping mode. 
For comparison and discussion, we also retrieved the 
{\it Spitzer} IRS observations of A\,58 obtained on 
2004 October 23 as part of the Program ID.\,3362 
(PI: A.\,Evans: AORKey: 10838272 and 10841600).

The low-resolution IRS observations were obtained with the short-low (SL) and long-low (LL) modules, covering together the wavelength range $\lambda=$5--38~$\mu$m. The IRS data also include observations obtained with the high-resolution modules short-high (SH) and long-high (LH), which have slits sizes of 4\farcs7$\times$11\farcs3 and 11\farcs1$\times$22\farcs3, respectively, positioned at the CSPNe. The spectral range covered by the high-resolution IRS observations is $\lambda=$10--37~$\mu$m. 
All IRS observations were processed using the CUbe Builder for IRS Spectra Maps
\citep[][]{Smith2007}. We note that the low-resolution spectra of A\,30 and A\,78 were analyzed by 
\citet{GarciaHernandez2013} in comparison with the extremely H-deficient R Coronae Borealis (RCB) stars to discuss the C-rich origins of these sources. Those authors were mainly interested in the broad 
dust features below 10~$\mu$m (the RCB stars generally display featureless 
spectra at longer wavelengths) and did not analyze the 
high-resolution {\it Spitzer} spectra.

Near-IR images of A\,30 and A\,78 were obtained with the NOTCam at the Nordic Optical Telescope (NOT). 
The observations were obtained on 2014 April 15--16 and August 9--10 through 
$J$ ($\lambda_\mathrm{c}=1.25 \mu$m, $\Delta\lambda=0.16 \mu$m), 
$H$ ($\lambda_\mathrm{c}=1.63\mu$m, $\Delta\lambda=0.29 \mu$m), and 
$K_\mathrm{s}$ ($\lambda_\mathrm{c}=2.14 \mu$m, $\Delta\lambda=0.28 \mu$m) filters. 
The total $J$, $H$ and $K$ exposure times for A\,30 were 120, 180 and 240~s and  144, 144 and 192~s for A\,78. All near-IR observations were reduced following standard {\sc iraf} routines \citep[][]{Tody1993}. The resultant $J$, $H$ and $K$ images of A\,30 and A\,78 are presented in the top panels of Figures~\ref{fig:A30} and \ref{fig:A78}, respectively.

Mid-IR Gran Telescopio de Canarias (GTC) CanariCam observations of
A\,78 were also obtained on 2014 Dec 12--13 and 2015 August 3 (PI:
M.A.\,Guerrero). Images were acquired using the Si3-9.8
($\lambda_\mathrm{c}$=9.8~$\mu$m, $\Delta\lambda=1.0~\mu$m), NeII-12.8
($\lambda_\mathrm{c}$=12.8~$\mu$m, $\Delta\lambda=0.2~\mu$m), and
NeII\_ref2-13.1 ($\lambda_\mathrm{c}$=13.1~$\mu$m,
$\Delta\lambda=0.2~\mu$m) filters\footnote{For a complete list of GTC
  CanariCam filters and their details see
  \url{http://www.gtc.iac.es/instruments/canaricam/canaricam.php\#Filters}.}
with total exposure times of 1500~s on each filter. The CanariCam
images were reduced following the {\sc RedCan} reduction pipeline
\citep[][]{GM2013}. It is important to notice that the CanariCam
observations cover a field of view of $26''\times19''$ and only
partially map the inner region of A\,78 (see bottom panels in
Fig.~\ref{fig:A78}).

For comparison with the IR images and further discussion we also
include in our analysis the optical (ground-based and {\it HST})
[O\,{\sc iii}] and X-ray {\it XMM-Newton} EPIC images of A\,30 and
A\,78 published by our team in previous works
\citep[see][]{Fang2014,Guerrero2012,Toala2015}. The [O\,{\sc iii}] narrow-band
images of A\,30 and A\,78 are presented in the left panels of Figure~\ref{fig:fig1}.

\begin{figure*}
\begin{center}
  \includegraphics[angle=0,width=\linewidth]{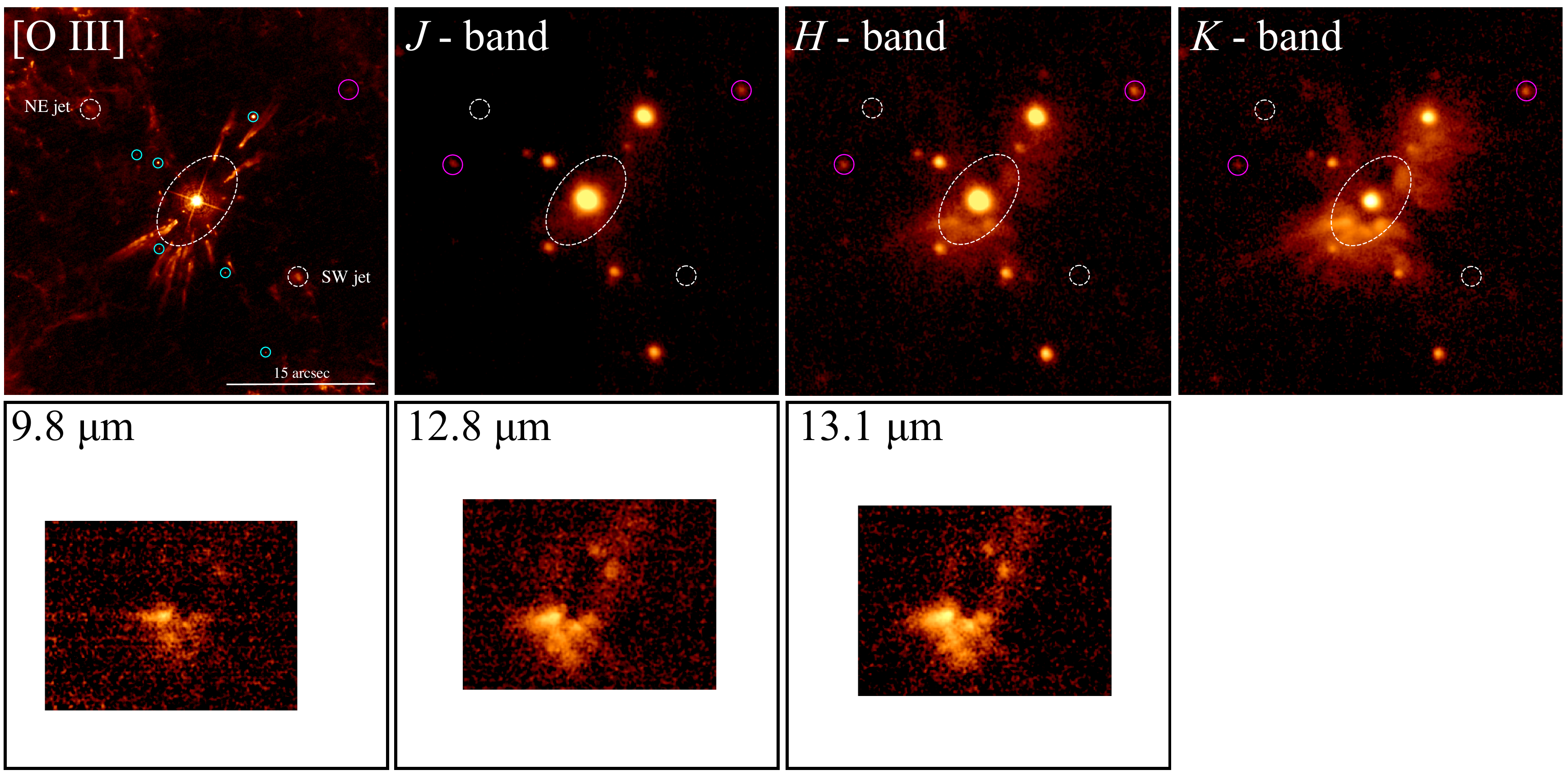}
\label{fig:A78}
\caption{
The inner region of A\,78 as seen in different wavelengths.  
The top panels show the {\it HST} [O\,{\sc iii}] narrow-band (leftmost panel) and NOTCam $J$, $H$ and $K$ images, whereas the bottom panels show GTC CanariCam images in different bands. 
The disk extension is shown with a dashed-line ellipse while the positions of the jet features are marked by dashed-line circles in the top panels. 
Background stars are shown with cyan circles.  
The position of clumps that appear in the NOTCam images are shown with a magenta circles. 
All panels correspond to the same FoV, but note that the CanariCam observations only map a region of $26''\times19''$ of the central area of A\,78. North is up, East to the left.}
\label{fig:A78}
\end{center}
\end{figure*}

\section{Spatial distribution of dust}

Figure~\ref{fig:fig1} shows the IR images of A\,30 and A\,78 in comparison with their optical [O\,{\sc iii}] images.  
The {\it Spitzer} IRAC and {\it WISE} images of A\,30 and A\,78 unveil in unprecedented detail the distribution of the hot dust around their CSPN. 
The dust extends up to distances $\gtrsim$40$''$ from the CSPNe with the brightest
region spatially coincident with the disrupted disk observed in
optical images \citep[see][and references therein]{Fang2014}. 
In the case of A\,30, a bipolar structure with PA=$-30^{\circ}$ can be easily
distinguished\footnote{The zero point of the PAs is defined by the north (PA=0$^\circ$) and increase counterclockwise.}. 
This is aligned with the bipolar jet seen in optical wavelengths, with NW and SE jet components located 6\farcs8 and 7\farcs8 from the star, but it seems to extend further out. 
The lower spatial resolution {\it WISE} images of A\,78 hinder a similar assessment. 

Colour-composite images combining the IR images of A\,30 and A\,78 with those of their optical [O\,{\sc iii}] and soft (0.3--0.5~keV) X-ray {\it XMM-Newton} images are presented in the right panels of Figure~\ref{fig:fig1}.  
These pictures show that the hot dust is not only spatially coincident with the hydrogen-poor knots and filaments, but also coexists with the X-ray-emitting material \citep[][]{Guerrero2012,Toala2015}.

\begin{figure}
\begin{center}
\includegraphics[angle=0,width=\linewidth]{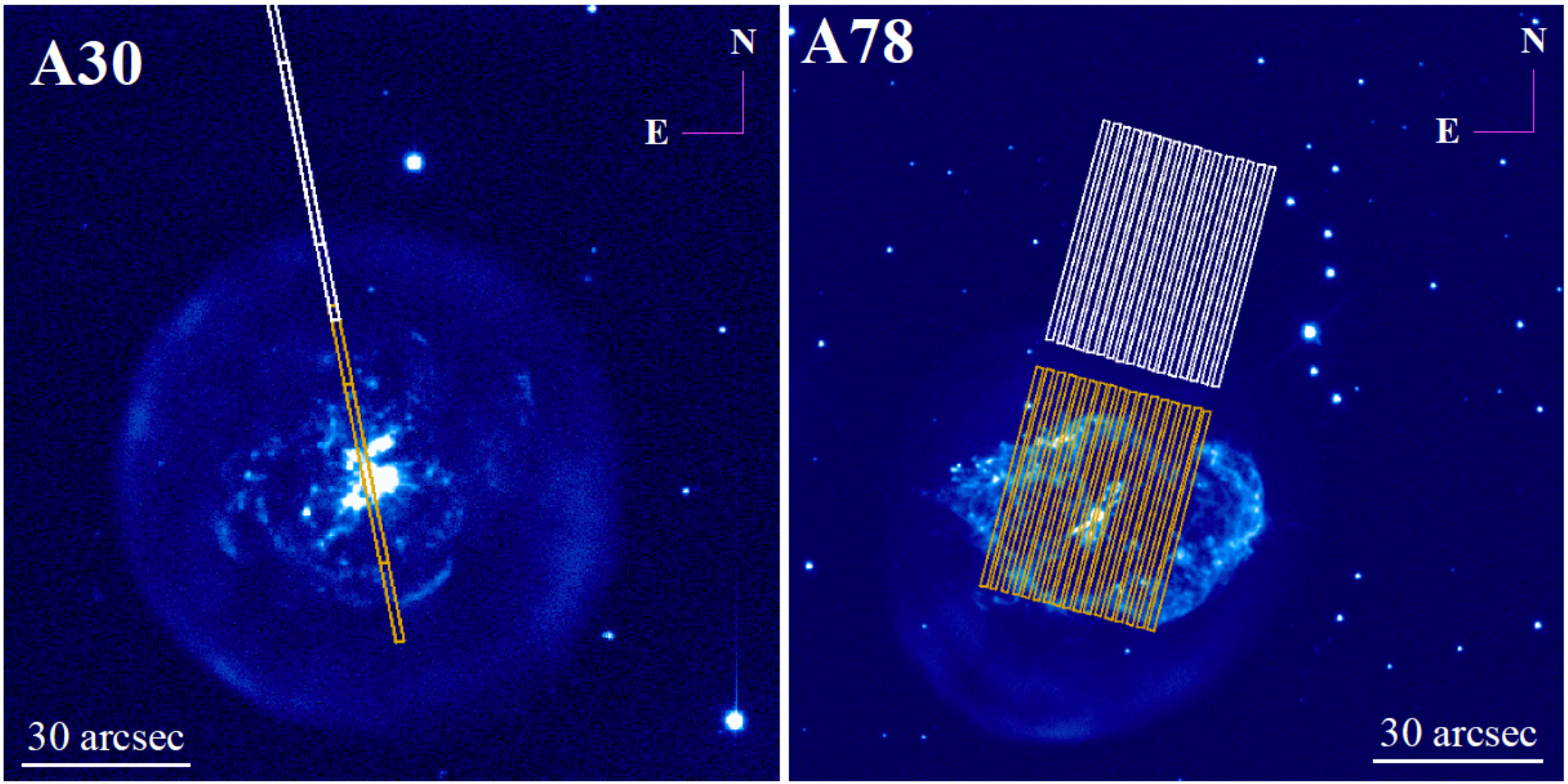}
\caption{Slit positions of the {\it Spitzer} IRS observations of A\,30 and A\,78. 
The orange rectangular regions show the positions of the short-low (SL) slits and the white rectangles those used for background subtraction.}
\label{fig:spitzer}
\end{center}
\end{figure}

Figures~\ref{fig:A30} and \ref{fig:A78} present near-IR images of the
central regions of A\,30 and A\,78, respectively, in comparison with
their corresponding {\it HST} [O\,{\sc iii}] images. The NOTCam $J$-band images show only emission from the CSPNe and background stars, although there is the marginal detection of extended emission in A\,78. Otherwise, the $H$- and $K$-band images clearly detect the emission from the disrupted disks of A\,30 and A\,78. The jets are undetected in the near-IR in agreement with previous reports  \citep[e.g.,][]{Dinerstein1984,Borkowski1994}, but they are detected in the \emph{Spitzer} IRAC images of A\,30 at longer wavelengths (see Fig.~\ref{fig:A30} bottom panels). An inspection of the images shows that the jet-like features extend 15.9$''$ towards the NW and 18.2$''$ to the SE from the CSPN. Their tips are coincident with hydrogen-deficient clumps identified in Figure~\ref{fig:A30} top left panel which are aligned with the knots identified as the NW and SE jet components. We note that the A\,78 $K$-band image is suggestive of emission located around the location of the jets (see Fig.~\ref{fig:A78} top right panel).

Figures~\ref{fig:A30} and \ref{fig:A78} also show that the clumps in the disrupted disks of A\,30 and A\,78 do not present uniform surface brightness. As for A\,30, the SW clumps are more IR bright than other clumps, whereas in A\,78 the brightest knots are located SE of the CSPN. This spatial distribution is similar to that presented in the [O\,{\sc iii}] {\it HST} images, as well as in the high spatial resolution {\it Chandra} X-ray image of A\,30, which shows an X-ray emission peak SW of the CSPN \citep[see figure~6 in][]{Guerrero2012}.

\section{Spectral properties of dust in A\,30 and A\,78}

\begin{figure*}
\begin{center}
  \includegraphics[angle=0,width=0.9\linewidth]{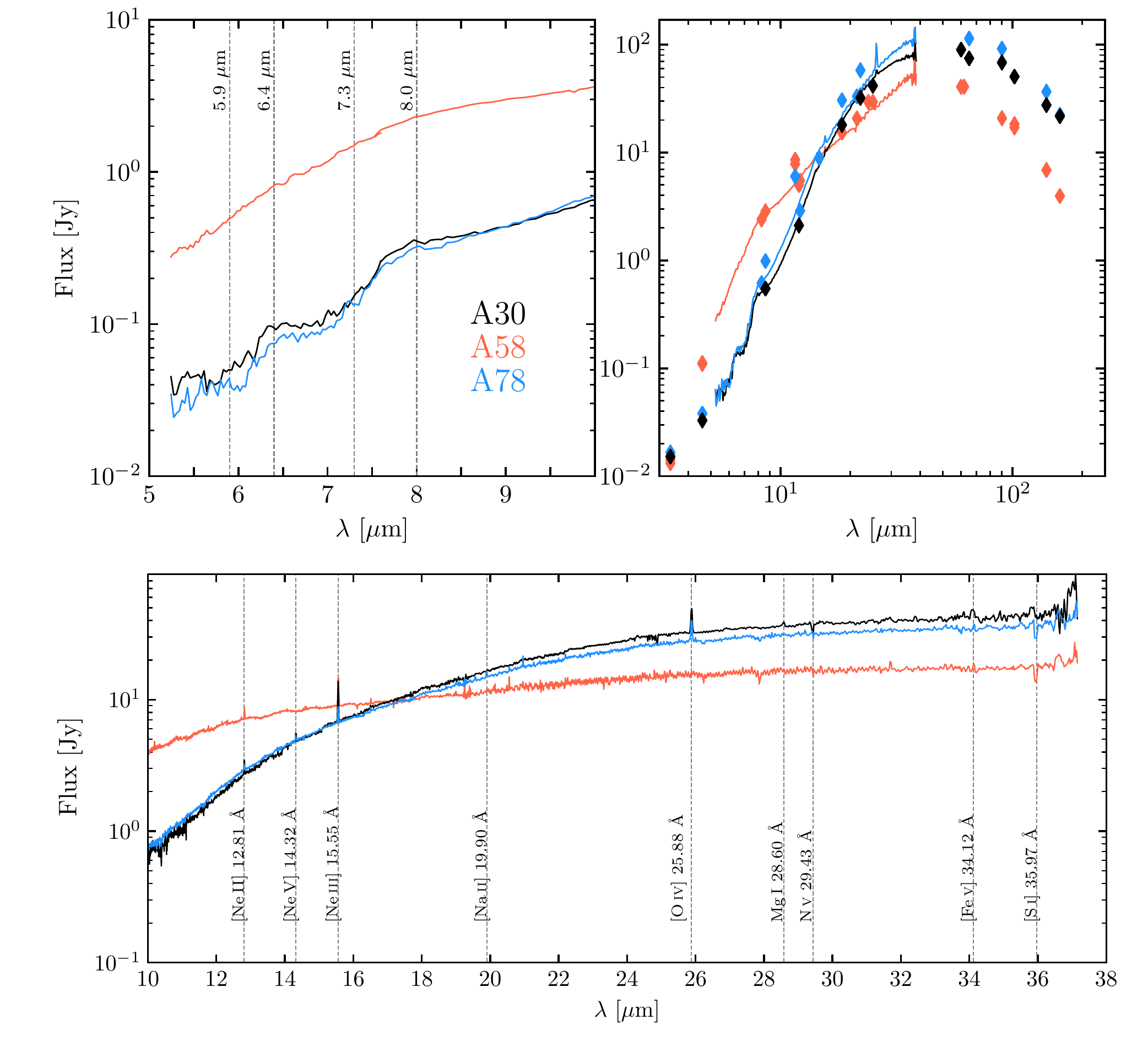}
\caption{
{\it Spitzer} IRS spectra of A\,30, A\,58 and A\,78. 
Top panels: low-resolution IRS spectra. 
The top-left panel shows the 5--10~$\mu$m range with the central wavelengths 
of the amorphous carbon features found by \citet{GarciaHernandez2013} shown 
with vertical dashed lines at 
6.4, 7.3 and 8.0~$\mu$m. 
The top-right panel shows the complete wavelength range with the corresponding 
photometry plotted with diamonds for each born-again PN. 
Bottom:  High-resolution IRS spectra. 
The most prominent atomic and ionic lines are marked.}
\label{fig:spec}
\end{center}
\end{figure*}

Figure~\ref{fig:spitzer} shows the slit positions of the low-resolution {\it Spitzer} IRS observations of A\,30 and A\,78.  
The background was selected from slits without any contribution from the H-deficient knots (shown as white slits of Figure~\ref{fig:spitzer}). 
The resultant background-subtracted spectra are shown in Figure~\ref{fig:spec}. 

The {\it Spitzer} IRS spectra of A30 and A78 are almost identical, which reflects the similar evolutionary stage and born-again origin discussed in previous works \citep[e.g.,][]{Fang2014,Toala2015}. The low-resolution IRS spectra of A\,30 and A\,78 presented in the top left panel of Figure~\ref{fig:spec} show a number of broad features, which were thoroughly discussed by \citet{GarciaHernandez2013}. Those authors compared the spectra of A\,30 and A\,78 with those of RCBs. After an appropriate continuum subtraction technique, \citet{GarciaHernandez2013} demonstrated that RCB objects exhibited the presence of broad features at 5.9, 6.4, 7.3 and 8.0~$\mu$m whilst the born-again PNe only present the latter three broad features \citep[see Figure~13 in][]{GarciaHernandez2013}. These authors concluded that the presence of such broad features can be attributed to amorphous carbon formed in H-deficient environments. 
Although subtle differences can be appreciated between A\,30 and A\,78, which  can be attributed to the signal-to-noise in the spectra, both sources exhibit the broad features with the 6.4 and 8.0~$\mu$m as the strongest and clearest features in Figure~\ref{fig:spec} top left panel.

\begin{figure}
\begin{center}
  \includegraphics[angle=0,width=0.92\linewidth]{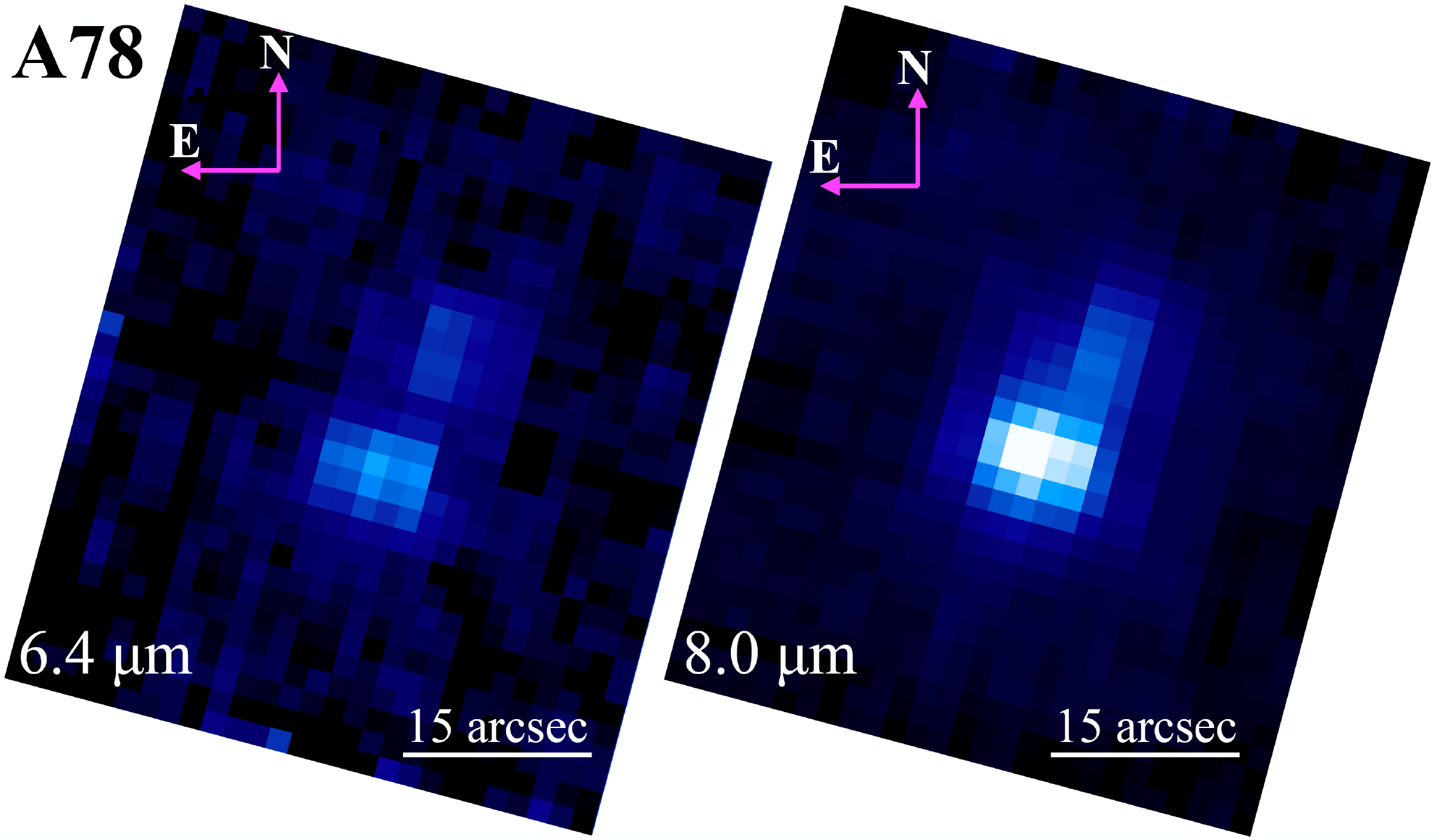}
\caption{{\it Spitzer} IRS spectral maps of A\,78 extracted at the 6.4 (left panel) 
and 8.0~$\mu$m (right panel) broad spectral features. 
The pixel size of the IRS maps is 1\farcs8.}
\label{fig:spitzer2}
\end{center}
\end{figure}

The IRS observations of A\,78 performed in mapping mode offer us the possibility to create spectral maps. 
Spectral maps of the 6.4 and 8.0~$\mu$m broad features are shown in Figure~\ref{fig:spitzer2}. 
We note that we also attempted to create images of the other broad features, but they resulted in low-quality images due to their lower signal-to-noise.
The images of the 6.4 and 8.0~$\mu$m broad features in A\,78 show that these carbon-rich bumps are mainly associated with the disk. 
Furthermore, the spectra of A\,78 show that the CanariCam images presented in the bottom panels of Figure~\ref{fig:A78} are not due to emission lines, but to dust continuum.

The bottom panel of Figure~\ref{fig:spec} presents the high-resolution IRS spectra.
The spectra do not present a large number of emission lines with the most prominent ones being those of [Ne\,{\sc iii}] 15.55 $\mu$m and [O\,{\sc iv}] 25.9 $\mu$m. 
Other less intense lines are those of [Ne\,{\sc v}] 14.32 $\mu$m, [Ne\,{\sc ii}] 12.81 $\mu$m and [Fe\,{\sc v}] 34.12 $\mu$m. 
Some absorption lines are also easily identified in the high-resolution spectra. 
In particular, we would like to note the N\,{\sc v} 29.43 $\mu$m absorption line which reflects the presence of gas with temperatures of $\sim10^{5}$~K, very likely the mixing region between the hot bubble ($\sim10^{6}$~K) and the nebular gas ($10^{4}$~K) \citep[see][and references therein]{Fang2016}. The N\,{\sc v} absorption line is detected with signal-to-noise larger than 30 for the three born-again PNe (see Appendix~\ref{sec:appA}).

For comparison with A\,30 and A\,78 we also show in Figure~\ref{fig:spec} the corresponding low- and high-resolution IRS spectra of the younger born-again PN A\,58. The IRS spectra of A\,58 are more intense for wavelengths below 15~$\mu$m and is below those of A\,30 and A\,78 for longer wavelengths. The high-resolution spectrum does not show the higher ionization Ne lines, but it exhibits the [Ne\,{\sc ii}]~12.81 $\mu$m line consistent with the lower $T_\mathrm{eff}$ of its CSPN compared to those of A\,30 and A\,78. 
It also shows several absorption features very likely due to molecular material present in the born-again ejecta of A\,58, similarly to what has been described for Sakurai's Object \citep[see][]{Evans2006,Evans2020}. 
We would like to note that the low-resolution spectrum of A\,58 also exhibits hints of the 8~$\mu$m broad feature better appreciated in the top right panel of Figure~\ref{fig:spec}.

Finally, we present in the top right panel of Figure~\ref{fig:spec} the available IR photometry for A\,30, A\,58 and A\,78 obtained from the NASA/IPAC IR Science Archive.

\section{Dust modelling}

Our observations have shown that the IR properties of A\,30 and
A\,78 are spatially and spectroscopically extremely similar. In
addition, previous multi-wavelength studies have shown that
these two born-again PNe share practically all their nebular
characteristics and stellar atmosphere analysis of their CSPNe
have also shown that they are {\it twins}
\citep[see][]{Guerrero2012,Toala2015}. 
In this section we will present a dust model tailored to A\,30 adopting a distance of $d$=2.4~kpc as estimated by \citet{BailerJones2018} using astrometric observations from {\it Gaia}\footnote{A subsequent paper addressing the dust and gas properties of A\,78, which is located at 1.5~kpc, using high-quality spectroscopic observations will be presented shortly (Montoro-Molina et al. in prep.)}.

Recent high-quality works have demonstrated that an appropriate description of dust in nebulae around evolved stars must also account for the behavior of gas in the calculations \citep[see, e.g.,][]{GLL2018,Rubio2020,JH2020}. 
Thus, in order to
peer into the dust properties of A\,30 and A\,78
we have used the photoionization code {\sc
  cloudy} \citep[version 17.0;][]{Ferland2017}, which  models 
  simultaneously the emission of ionized gas and
dust as a consequence of the presence of a radiation
field. By coupling the {\sc pyCloudy} libraries \citep{Morisset2013} we are able to 
produce synthetic observations to be directly compared to the real ones.

\begin{figure}
\begin{center}
\includegraphics[angle=0,width=\linewidth]{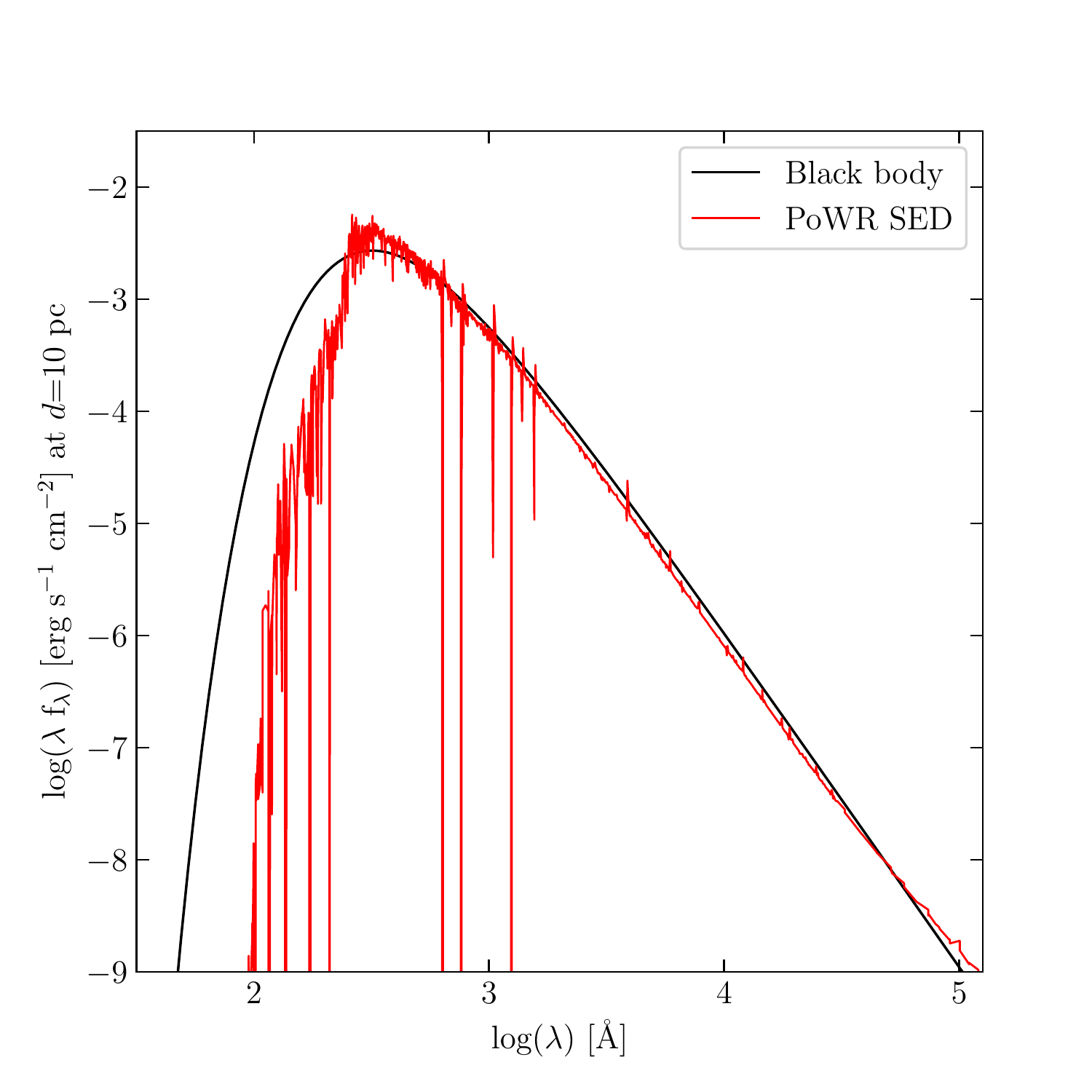}
\caption{Comparison between the best-fit model to the stellar
  atmosphere of A\,30 obtained with the PoWR code and a black body
  emission model with a temperature of 115kK.}
\label{fig:SED}
\end{center}
\end{figure}

{\sc cloudy} requires several user-defined inputs: 
i) the incident spectrum, 
ii) the chemical abundances,
iii) the density distribution and geometry, and 
iv) the dust physical properties (size and chemical composition) and spatial distribution. 
Our group has modeled the optical and
UV observations of the CSPN of A\,30 using the state-of-the-art
stellar atmosphere code
PoWR
\citep{Grafener2002,Hamann2004}\footnote{\url{http://www.astro.physik.uni-potsdam.de/~wrh/PoWR/powrgrid1.php}}. 
The best model for the stellar atmosphere of the CSPN of A\,30 has log($L$/L$_\odot$)=3.78
and $T_\mathrm{eff}$=115~kK \citep[see the details of the model in][]{Guerrero2012}. 
The synthetic
stellar atmosphere model obtained from PoWR 
will be used in {\sc cloudy} as the incident spectrum. 
This is compared to a black body emission model with the same temperature in Figure~\ref{fig:SED}, revealing the noticeable opacity of the dense atmosphere of Wolf-Rayet-type stars to hard photons.

There are not many  studies of the physical properties and chemical abundances of the hydrogen-deficient clumps in A\,30 in the
literature \citep{Jacoby1983,Guerrero1996,Wesson2003}. Here we will adopt the He, O, N and Ne abundances of knot~1 determined by \citet{Guerrero1996} complemented with the C/O=0.3 by mass reported by \citet{Wesson2003}. Other elemental abundances were set to their solar values \citep{Wilms2000}.

\begin{figure}
\begin{center}
\includegraphics[angle=0,width=\linewidth]{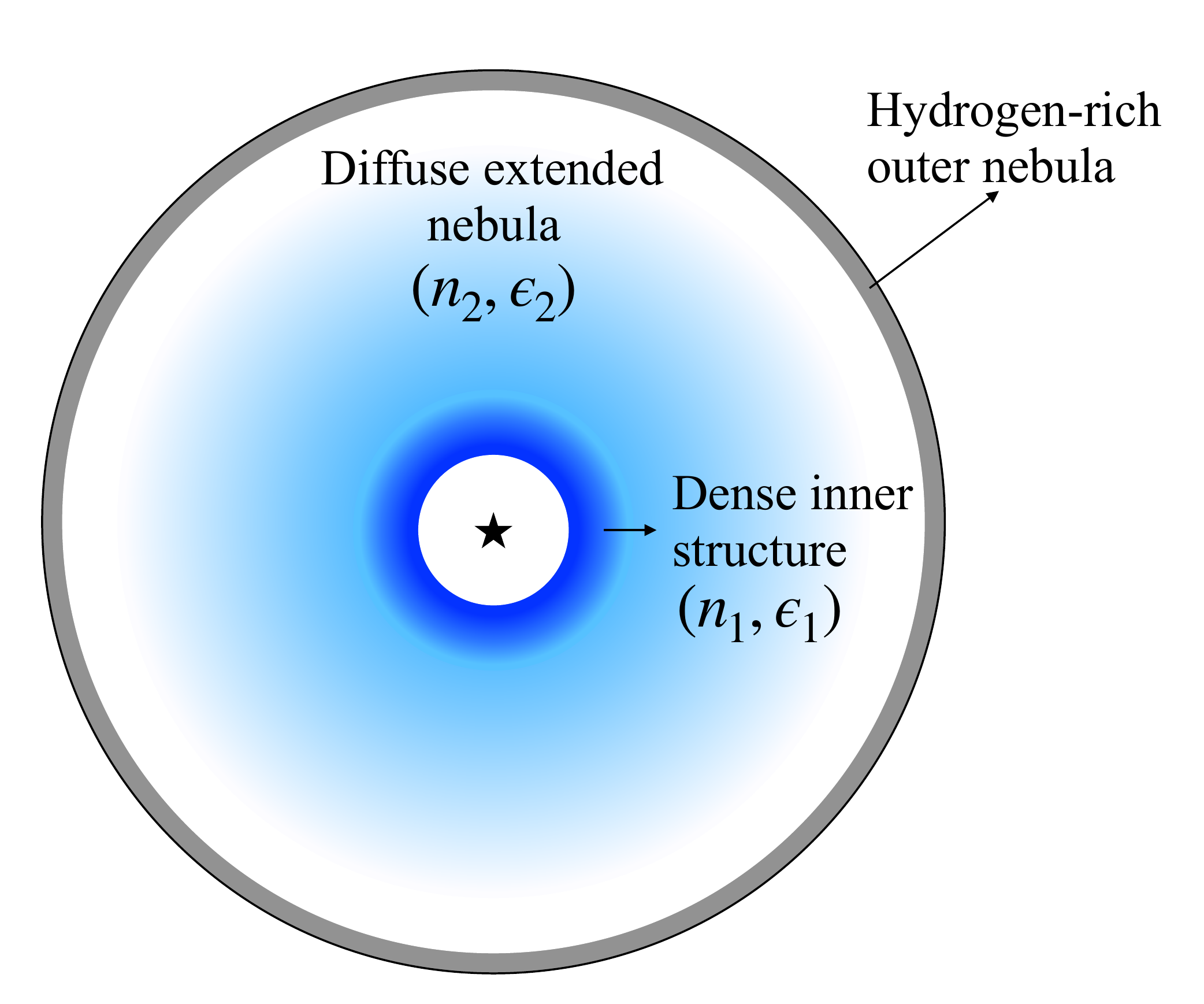}
\caption{Two-shell distribution model adopted in our {\sc cloudy} models.}
\label{fig:esquema}
\end{center}
\end{figure}

\begin{figure*}
\begin{center}
  \includegraphics[angle=0,width=0.93\linewidth]{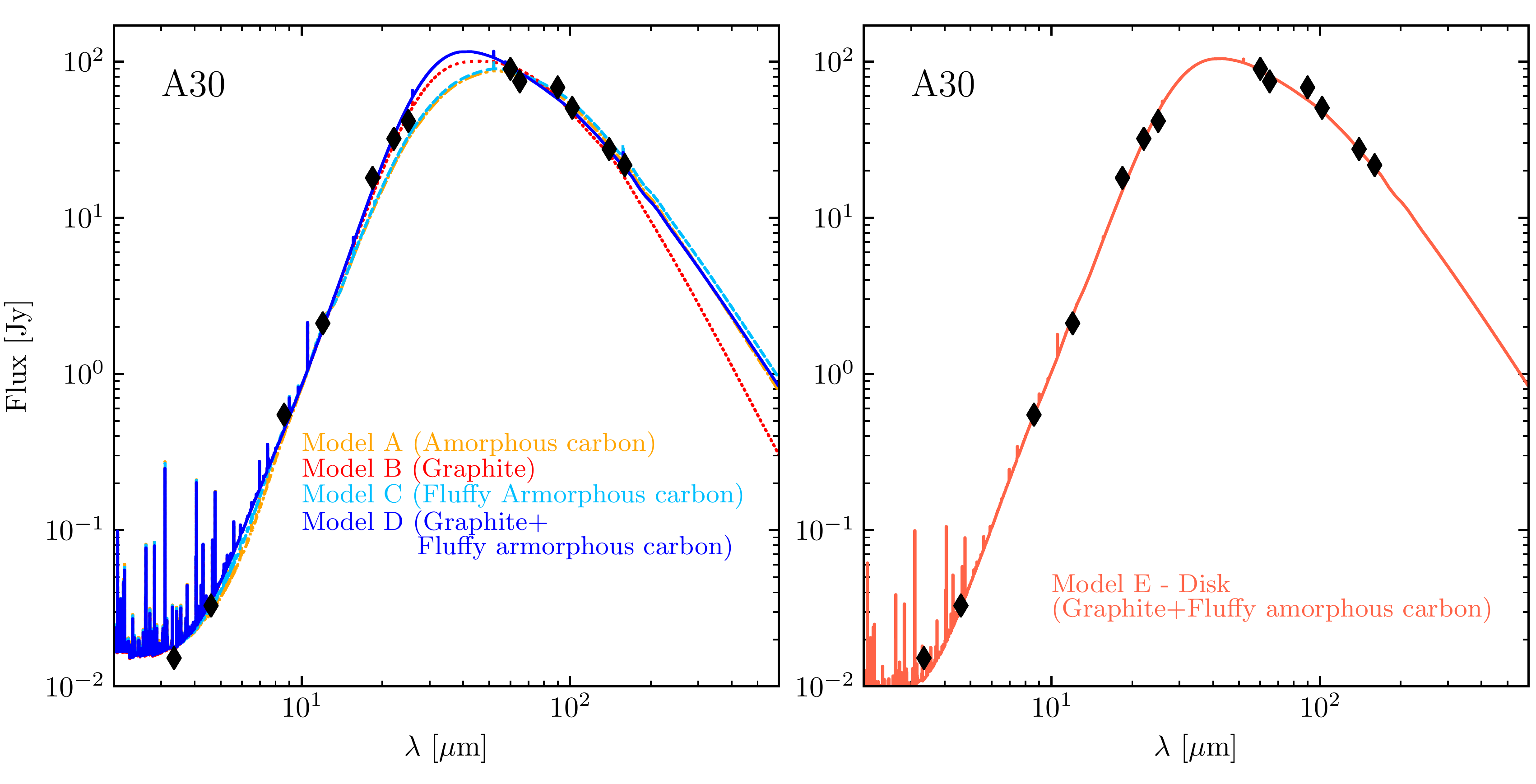}
  \caption{Comparison between the observed IR SED of A\,30 (black diamonds) 
    with our {\sc cloudy} models. Different colours and line styles 
    shown the spectrum from
    models as labeled on the panel. The right panel shows our best model
    taking into account a disk morphology for the inner region.}
\label{fig:models}
\end{center}
\end{figure*}

For comparison with previous models of the dust in A\,30, we started by 
taking into account two structures to
model the nebular and IR properties of this born-again PN. A dense shell
structure close to the CSPN with inner and outer radii of
$r_\mathrm{in}$=8.6$''$ and $r_\mathrm{mid}=$15$''$, respectively. These radii have
been obtained by averaging the dense, hydrogen-deficient clumps
distances to the CSPN of A\,30. A secondary
shell structure was defined with an inner radius of
$r_\mathrm{mid}$=15$''$ and an outer radius $r_\mathrm{out}$ which
will be kept as a free parameter in our model, but a maximum limit of
60$''$ is predefined by the optical radius of the hydrogen-rich PN. 
The total number densities and filling factors of the two shells are denoted
as $n_1$, $n_2$, $\epsilon_1$ and $\epsilon_2$, respectively. For both shells we
adopted density distributions with a dependence with radius. For example, for
the inner shell we define its density distribution as
\begin{equation}
  n(r) = n_1 \left( \frac{r}{r_1} \right)^{-\alpha},
\end{equation}
\noindent where $n_{1}$ is the total number density of the gas at its inner face located at $r_{1}$. A similar equation is defined for the outer shell. Figure~\ref{fig:esquema} presents a schematic view of this configuration.

{\sc cloudy} give us the possibility to include in our calculations 
different built-in dust species 
\citep[silicates and carbonaceous;][]{Ferland2017}. 
According to the evolution of the born-again event, we will
include carbon-rich dust. We will present a model including amorphous
carbon, but we will also compare with models including graphite\footnote{We note that \citet{Borkowski1994} rejected the presence of graphite in A\,30, however, laboratory experiments have demonstrated the graphitization of amorphous carbon material by the presence of a radiation field \citep[see, e.g.,][and references therein]{Ogmen1988,Kim2016}.} and
fluffy grains (porous aggregates of smaller grains). 
\citet{Borkowski1994} suggested that a dust size distribution with a slope of $-$3
described better the IR SED of A\,30 \citep[in contrast to the $a^{-3.5}$ suggested by][for the ISM]{Mathis1977}. 
Following these authors, we fixed the slope of the dust size distribution to $-3$ in all the
following calculations, but we change the limiting values of the dust sizes in different models. We start our model by using grain sizes
between 0.0007 and 0.25~$\mu$m with a power-law
distribution of radii of $\propto a^{-3}$. However, in order to 
improve our models we had to define dust populations 
with small ($a_\mathrm{small}$) and big sizes ($a_\mathrm{big}$).
Each distribution of sizes is divided into 10 bins. 

Once the model is computed, synthetic nebular observations 
will be obtained using {\sc pyCloudy} for a slit 1.5 arcsec 
in width and 3.7 arcmin in length, similar to the one used 
in \citet{Guerrero1996}.

\subsection{Model exploration}

We remark that {\sc cloudy} not only takes into account the contribution from dust, but also computes the properties of the gaseous component. 
That is, the 1-5~$\mu$m wavelength range includes not only the contribution from small grains \citep[as discusses in][]{Borkowski1994}, but also the contribution from emission lines and continua from the ionized gas (see Fig.~\ref{fig:models}). 
The combination of the density and small grain size distribution need to be accounted for simultaneously. 
However, one can restrict the contribution from the ionized gas by trying to fit the H$\alpha$ flux and the electron temperature $T_\mathrm{e}$([O\,{\sc iii}]) estimated using the [O\,{\sc iii}] emission lines. 
\citet{Guerrero1996} reported log(H$\alpha$)$\approx -15$ 
and $T_\mathrm{e}$([O\,{\sc iii}])=14,000~K, whilst \citet{Wesson2003} report 
$T_\mathrm{e}$([O\,{\sc iii}])$\lesssim$18,000~K using different emission lines. 
We found that for the inner shell $n_0$=2100~cm$^{-3}$ and
$\epsilon=7.7\times10^{-4}$ closely reproduce the H$\alpha$ flux and electron 
temperatures. This electron
density agrees with that reported in
\citet{Guerrero1996} for knot~1 (2100~cm$^{-3}$). 
Thus, for the four models described in the following
we fixed $n_\mathrm{0}$.

\begin{table*}
\centering
\setlength{\columnwidth}{0.1\columnwidth}
\setlength{\tabcolsep}{1\tabcolsep}
\caption{Details of our {\sc cloudy} models of the nebular and dust in
  born-again PNe.}
\begin{tabular}{lccccc}
\hline
\multicolumn{1}{l}{Parameter} &
\multicolumn{1}{c}{Model~A}&
\multicolumn{1}{c}{Model~B} & 
\multicolumn{1}{c}{Model~C} & 
\multicolumn{1}{c}{Model~D} &
\multicolumn{1}{c}{Model~E} \\
\multicolumn{1}{l}{} &
\multicolumn{1}{c}{(Amorphous}&
\multicolumn{1}{c}{(Graphite)} & 
\multicolumn{1}{c}{(Fluffy amorphous} & 
\multicolumn{1}{c}{(Graphphite$+$Fluffy} &
\multicolumn{1}{c}{(Graphite+Fluffy}  \\ 
\multicolumn{1}{c}{}  & 
\multicolumn{1}{c}{carbon)} &
\multicolumn{1}{c}{} &
\multicolumn{1}{c}{carbon)} &
\multicolumn{1}{c}{amorphous carbon)} & 
\multicolumn{1}{c}{amorphous carbon)} \\
\hline
Inner structure                                & shell  & shell  & shell  & shell & disk  \\
$n_{1}$ (cm$^{-3}$)                            & 2100   & 2100   & 2100   & 2100  & 2100  \\
$\alpha$                                       &  3     &  3     &  3     &  3     &  3  \\
$\epsilon$ $(\times10^{-4})$                   & 7.7    & 7.7    & 7.7    & 7.7    & 7.7 \\
$r_\mathrm{in}$ ($''$)                         & 8.6    & 8.6    & 8.6    & 8.6    & 8.6 \\
$r_\mathrm{mid}$ ($''$)                        & 15     & 15     & 15     & 15     & 15  \\
Amorphous carbon size distribution             & 0.0008--0.03 & \dots       & 0.0011--0.02  &  \dots  &  0.0011--0.02\\
Graphite size distribution                     & \dots  & 0.001--0.03 & \dots   & 0.001--0.03 & 0.001--0.03 \\
M$_\mathrm{gas}$ ($\times10^{-3}$~M$_{\odot}$) & 10.06  & 10.06  & 10.06  & 10.06  & 2.85\\
M$_\mathrm{dust}$($\times10^{-3}$~M$_{\odot}$) & 0.036   & 0.23  & 0.029  & 0.23   & 0.10\\
gas-to-dust                                    & 278     & 45    &353     & 45     & 25\\
\hline
Outer Shell                                    &        &        &        &        &      \\
$n_2$ (cm$^{-3}$)                              & 55     & 55     & 55     & 55     &  55  \\
$\alpha$                                       &  3     &  3     &  3     &  3     &   3  \\
$\epsilon (\times10^{-4})$                     &  4     & 4      & 4      & 4      &   4  \\
$r_\mathrm{mid}$ ($''$)                        & 15     & 15     & 15     & 15     &   15 \\
$r_\mathrm{out}$ ($''$)                        & 50     & 50     & 50     & 50     &   50 \\
Amorphous carbon size distribution             & 0.04--0.10 & \dots  & 0.025--0.10 & 0.06--0.20 &  0.06--0.20 \\
Graphite size distribution                     & \dots  & 0.30--0.60 & \dots  & \dots  & \dots  \\
M$_\mathrm{gas}$ ($\times10^{-3}$~M$_{\odot}$) & 1.56   & 1.56   & 1.56   & 1.56   &  1.56\\
M$_\mathrm{dust}$($\times10^{-3}$~M$_{\odot}$) & 6.30   & 19.6   & 2.94   & 3.10   &  3.10\\
Gas-to-dust                                    & 0.25   &  0.07  & 0.53   & 0.50   &  0.50\\ 
\hline
$M_\mathrm{TOT,gas}$ ($\times10^{-3}$~M$_{\odot}$) & 11.62 & 11.62  & 11.62  & 11.62 &  4.41  \\
$M_\mathrm{TOT,dust}$($\times10^{-3}$~M$_{\odot}$) & 6.34  & 19.83  & 2.97   & 3.32  &  3.20  \\
\hline
log($F_{{\rm H}\alpha}$) (erg~cm$^{-2}$~s$^{-1}$)                & $-$15.03  & $-$15.06 & $-$15.04 & $-$15.06 & $-$15.09 \\
$T_\mathrm{e}$([O\,{\sc iii}]) (K)            & 21,400    & 16,600   & 22,400   & 18,400   & 18,500   \\ 
 \hline
 Observed                                      & & & & \\
 log($F$(H$\alpha$))                           & $-$15.04$\pm0.09$ & & &  \\
 $T_\mathrm{e}$([O\,{\sc iii}]) (K)            & 14,000--18,000   & & &  \\
\hline
\end{tabular}
\label{tab:models}
\vspace{0.15cm}
\end{table*}

We started by adopting amorphous carbon with sizes between
0.0007 and 0.25~$\mu$m prescribed by \citet{Borkowski1994}, but no successful 
fit could be achieved. 
In short, dust of these sizes produce considerable emission in the 1--5~$\mu$m 
that exceeds the observed emission of A\,30. 
This simple exercise showed that
grains as small as 0.0007~$\mu$m cannot be used to reproduce the IR properties of 
A\,30 in conjunction with its optical properties.

In order to appropriately model the SED of A\,30 including dust and gas, 
we need to fit the size range of our dust distribution. 
Our first model, Model~A, required two population of amorphous carbon grains, 
a population of small grains with sizes of 0.0008--0.03~$\mu$m 
present only in the inner shell and larger grains with sizes between 
0.04--0.10~$\mu$m present only in the outer region. 
Details of the model are listed in Table~\ref{tab:models}. 
Model~A predicts a total dust mass of 
$M_\mathrm{dust,A}=6.34\times10^{-3}$~M$_\odot$, with 99.4\% corresponding to dust in 
the outer shell. The total predicted mass of gas, which is computed from the 
H$\alpha$ flux, 
is $M_\mathrm{gas,A}=1.16\times10^{-2}$~M$_\odot$ with 87\% of this retained in the inner
shell. Model~A is compared to the observed SED in the left panel 
of Figure~\ref{fig:models} (yellow line). 
This figure shows that Model~A
does a good job fitting the IR SED of A\,30 with a predicted 
peak at 50--60~$\mu$m, but there is certain deviation from the
photometry for wavelengths between 15 and 30~$\mu$m. \citet{GLL2018} discussed a
similar problem for the normal H- and C-rich PN IC\,418 and attribute it  
to the presence of hydrogenated carbonaceous species not included in {\sc cloudy} 
(see Discussion section). However, we note that the specific cases 
of A\,30 and A\,78 studied here are different as no hydrogenated dust 
species are expected (see previous section). 

A second model was performed with the same density distribution but 
including only graphite in both shells, Model~B. 
This model required that the size ranges of both the smaller and larger 
grains be shifted to larger values, of 0.001--0.03~$\mu$m and much bigger 
large grains of 0.30--0.60~$\mu$m.
Model~B resulted in a 
larger total dust mass of $M_\mathrm{dust,B}=1.98\times10^{-2}$~M$_\odot$.
Around 99\% of the total dust is also required to be present in the outer shell, similarly
to Model~A. Model~B is shown as a red line in Figure~\ref{fig:models} left panel. 
This model is better than Model~A fitting the IR SED 
around 15--30~$\mu$m, but some emission is lacking at longer wavelengths. 
Furthermore, its SED peak appears displaced towards shorter wavelengths, peaking at $\sim$40~$\mu$m, when compared to Model~A. 
Details of the model are listed in Table~\ref{tab:models} labeled as Model~B.

Model~C is similar to Model~A but adopting {\it fluffy} amorphous carbon grains.
We used {\sc cloudy} to construct distributions of amorphous carbon 
with empty fractions 
of 50\%. To produce a good fit to the IR SED, this model 
required dust populations of small and big grans with sizes of 0.0011--0.02~$\mu$m
and 0.025--0.10~$\mu$m, respectively. 
The total dust mass resulted in $M_\mathrm{dust,C}=2.97\times10^{-3}$~M$_\odot$ 
with 99\% of the dust in the
outer shell. Figure~\ref{fig:models} left panel shows that the synthetic SED of 
Model~C is very similar to that of Model~A 
although it resulted in only 15\% of the total dust mass because these lighter 
fluffy grains have higher absorption cross-sections. Similarly to Model A, 
Model C peaks between 50--60~$\mu$m and has trouble fitting the SED around 15--30~$\mu$m.

Model~D was performed taking into account a combination of graphite 
and fluffy amorphous carbon dust. This models required dust populations of small
and big grains of 0.001--0.03~$\mu$m and 0.06--0.20~$\mu$m, respectively, with total 
dust mass of $M_\mathrm{dust,D}=3.32\times10^{-3}$~M$_\odot$. This model requires 93\% of
the total dust mass to be present in the outer shell. Model~D does a better
job than all previously presented models, fitting simultaneously most of the photometric
measurements presented in Figure~\ref{fig:models}.

\begin{figure*}
\begin{center}
  \includegraphics[angle=0,width=0.95\linewidth]{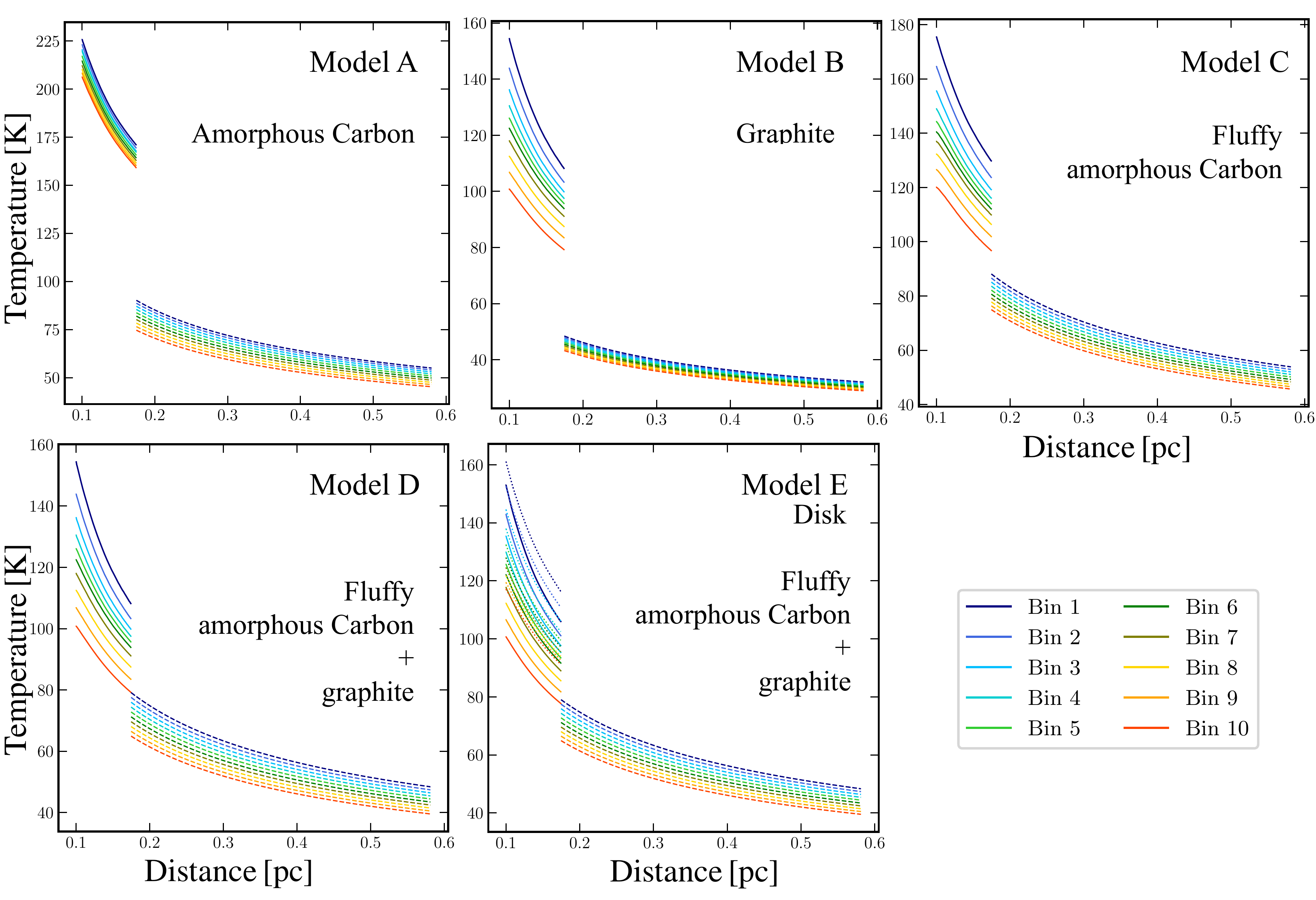}
  \caption{Temperature distribution of the different dust grain sizes obtained from the 
  models presented in Section~5 (see Table~\ref{tab:models}) 
  as a function of distance to the CSPN. 
  Each colour represents different bins in the
  dust size distributions. The solid (dashed) lines show the temperature of the
  small (large) grains. Note the differences in temperature scale between 
  all panels. In particular, in the Model~E panel the dotted lines show 
  the temperature for the graphite grains in the inner disk.}
\label{fig:dust_temp}
\end{center}
\end{figure*}

Following our findings the model that best reproduces the observed IR SED of A\,30 
in the mid- and far-IR includes the contribution from fluffy amorphous carbon and graphite (Model D). 
In order to create a more realistic model of the distribution of dust and gas in A\,30, 
we redefined the inner structure of the model. 
Instead of adopting a spherical shell, we now adopted a disk (torus) structure. 
The inner and outer radii are taken to be the same as for the inner shell in previous models, 
8.6 and 15 arcsec, respectively, with a thickness of 6.4 arcsec. 
All of the parameters of the outer shell  are the same as those of Model~D, however, 
we now include a combination of graphite and fluffy amorphous carbon in the inner 
disk structure and only fluffy amorphous carbon in the outer shell.
Details of this model are listed in Table~\ref{tab:models} as Model~E.

Model~E resulted in total mass of gas 
$M_\mathrm{gas,E}=(4.41^{+0.55}_{-0.14})\times10^{-3}$~M$_\odot$, with 65\% 
of the total mass in the inner disk. 
We note that the errors in the estimated ionized mass were now computed 
taking into account the errors in the H$\alpha$ flux (see bottom rows of 
Table~\ref{tab:models}). The total mass of dust resulted in $M_\mathrm{dust,E}=3.20\times10^{-3}$~M$_\odot$ 
with 95\% of the dust to be in the outer shell. This model is compared to the observed
SED in the right panel of Figure~\ref{fig:models}. As a result of the smaller mass 
of the disk compared to a shell, the contribution of the emission lines in the 1--5~$\mu$m
has diminished compared to the other models. We would like to note that whereas the
mass of the gas is constrained by the $F$(H$\alpha$), the total mass of dust depends
strongly on how empty the fluffy grains are. We reckon that an empty fraction of 50\% 
is an arbitrary value and thus, computed a range of models covering different fractions. 
Taking into account fluffy amorphous carbon 
grains with 0 and 90\% empty fractions we estimate 
that the dust mass of a disk model is 
$M_\mathrm{dust,E}=(3.20^{+3.21}_{-2.06})\times10^{-3}$~M$_\odot$, which 
still reproduce the IR SED.

In Figure~\ref{fig:dust_temp} we present the temperature distribution of the 5 
models presented in this section as a function of distance from the CSPN. 
All of our models predict dust temperatures in the 50--230~K range. In particular, 
Model~A accounting only for amorphous carbon has the highest dust temperatures 
in the inner shell with $T_\mathrm{dust}=$150--225~K. In all models, 
the large grains have temperatures between 
40 and 100~K.

\section{DISCUSSION}

Previous IR studies of A\,30 and A\,78 did not have the spatial resolution 
nor the FoV necessary to unveil the extension of the dust in these two 
born-again PNe. The analysis of mid-IR {\it Spitzer} IRAC and {\it WISE} images 
have shown that dust is present (at least) to radii $\sim40''$. 
The collection of IR images presented here have shown that the disrupted 
disk is brighter at shorter wavelengths while the surrounding clumps and 
filaments increase their brightness towards longer wavelengths, suggesting 
that small hotter grains are located close to the CSPNe.

The {\it Spitzer} IRS observations of A\,30 and A\,78 presented here
exhibit the first imaging evidence of amorphous carbon in these PNe. 
The broad features below 10~$\mu$m correspond to amorphous carbon 
formed in H-deficient environments \citep[e.g.,][]{GarciaHernandez2013}
and are mapped with the spectral mapping capabilities of the {\it Spitzer} 
IRS observations of A\,78, mainly located in the inner disk.
This situation is interesting and impacts even the discussion 
of the X-ray emission in born-again PNe. \citet{Guerrero2012} and 
\citet{Toala2015} found that the temperature of the X-ray-emitting 
gas within A\,30 and A\,78 are the softest among PNe. Although 
hydrodynamic mixing and/or thermal conduction is usually invoked 
to explain soft X-ray temperatures, born-again PNe are extremely 
soft sources ($T_\mathrm{X} \lesssim$10$^{6}$~K). \citet{Guerrero2012} 
found that the maximum in the X-ray emission
is coincident with the brightest H-deficient clump detected in the 
[O\,{\sc iii}] {\it HST} images. Accordingly, they 
suggested that charge exchange reaction processes
could be occurring on the surface of the C-rich dust 
\citep[see][]{Dennerl2010,Dennerl1997}
helping explaining the dominant C\,{\sc iv} emission line 
in the X-ray spectra.
We suggest that heavy ions from the stellar wind of the CSPN 
of A\,30 and A\,78 take electrons from the neutral amorphous 
carbon material and produce the soft X-ray 
emission.

To our knowledge, 
there is no further evidence of other molecules in A\,30 and A\,78
at IR wavelengths nor sub-millimeter observations, although we note that 
 complex molecules, such as hydrogenated 
and nitrogenated amorphous carbon (among others), have been reported 
to be present in Sakurai's Object 
\citep[see][]{Clayton1997,Evans2006,Evans2020,Eyres1998,Tafoya2017,Pavlenko2004}
and CO has been detected in A\,58 \citep{Tafoya2017}.

The analysis presented here suggests that modeling the IR SED of A\,30 and A\,78 
with purely amorphous carbon is a simplification and that other carbonaceous dust 
species must be included in the calculations. Following previous 
studies of the IR emission of born-again PNe, we started our {\sc cloudy} 
models adopting dust composed only by amorphous carbon. 
Small grain sizes as those suggested by \citet{Borkowski1994} produce an
excess in the near-IR when including the contribution of gas. UV photons 
ionize the gas and heat up the dust grains, that is, the resultant properties
of the IR emission are different as in the absence of the former.

The purely amorphous carbon case (Model~A) with a small grain size 
of 0.0008~$\mu$m has problems in fitting the IR SED between 15--30~$\mu$m. 
As mentioned before, \citet{GLL2018} modeled the IR properties of the H- 
and C-rich PN IC\,418 and argued that the emission in 
this range could be due to unknown carbon-rich hydrogenated 
species currently not included in {\sc cloudy}. For example, 
\citet{Kwok2011} showed that some broad features in this wavelength range 
can be explained by 
the presence of amorphous organic solids with a mixed 
aromatic–aliphatic structures and \citet{Grishko2001} showed that 
hydrogenated amorphous carbon produces broad features in the 20--30~$\mu$m
wavelength range. However, the specific case of born-again PNe would require
amorphous carbonaceous solids with little or no H. 
The inclusion of graphite (pure carbon solids) in the inner disk can help us addressing this issue.

Previous estimates of the dust temperature in born-again PNe differed considerably depending on the IR range modeled, with dust temperatures of 1000~K for near-IR $\sim$2--3~$\mu$m measurements \citep{Cohen1977}, but around 145~K with a marginal decrease of temperature with distance from the CSPN for mid-IR 10 and 20~$\mu$m observations \citep{Dinerstein1984}.  
We demonstrated here that an agreement is only possible if the contribution from the ionized gas is considered. 
Our test models resulted in dust temperatures between 40 and 225~K, with different ranges depending on the dust species included in the calculation, but a clear gradient of declining temperature towards the outer regions (see Fig.~\ref{fig:dust_temp}). 
Our disk model (Model~E) required small grains with temperatures in the $\sim80-160$~K and large grains in the 40--80~K temperature range. 
If one takes into account the dust temperatures estimated for Sakurai's Object (1000~K) and A\,58 (200--800~K), our results imply that the dust temperature decreases with time, with the dust in the most evolved born-again PNe (A\,30 and A\,78) having the lowest temperature.

Our best model that reproduces the nebular and IR properties of A\,30, Model~E, 
considers an inner disk-like distribution of matter composed of graphite and fluffy amorphous carbon grains. It predicts a total mass of dust of $M_\mathrm{dust}=(3.20^{+3.21}_{-2.06})\times10^{-3}$~M$_\odot$, which is very similar to the values of [2--5]$\times10^{-3}$~M$_{\odot}$ reported in previous works \citep[see][]{Moseley1980,Borkowski1994}. 
We reckon that modeling the amorphous carbon dust with fluffy grains 
might be more realistic compared to the usually adopted solid spheres.
Amorphous carbon dust are structured in chain-like distributions 
\citep{Rotundi1998}, whilst the optical and IR properties of graphite 
are less sensitive to shape \citep[see][]{Rai2017}.
However, the exact empty fraction is unknown and limits the capability of the models to determine a more accurate $M_\mathrm{dust}$.
To illustrate this, the range of dust mass presented here has been computed by a adopting 0\% and 90\% empty fractions for the amorphous carbon particles. Emptier dust particles include less dust mass but are still able to reproduce the IR SED because their absorption cross-sections are higher.

The total mass of gas predicted by Model~E is
$M_\mathrm{gas}=(4.41^{+0.55}_{-0.14})\times10^{-3}$~M$_\odot$. 
Then, the total mass ejected during the VLTP can be estimated to be
\begin{equation}
M_\mathrm{VLTP} = M_\mathrm{dust} + M_\mathrm{gas} = (7.61^{+3.76}_{-2.20})\times10^{-3} \mathrm{M}_\odot. 
\end{equation}
\noindent If the born-again event lasted 20--100~yr, as predicted by stellar evolution models \citep[see][]{MM2006}, the mass-loss rate during the VLTP would be $\dot{M}_\mathrm{VLTP}\approx[5-60]\times10^{-5}$~M$_{\odot}$~yr$^{-1}$, although due to the clumpy nature of the amorphous carbon dust, we favor the lower limits of $M_\mathrm{VLTP}$ and $\dot{M}_\mathrm{VLTP}$. 
Even then, this value is at the top of the reported mass-loss rates of AGB stars in the literature \citep[see][and references therein]{Ramstedt2020}, but we note that the origin of the VLTP is explosive. 
This is the first time that an estimation of the mass lost during the VLTP is attempted for A\,30. 
Interestingly, \citet{Guerrero2018} presented stellar evolution models to try to unveil the evolutive path of the born-again PNe HuBi\,1 and found that a
mass-loss rate during the born-again event of $7.6\times10^{-5}$~M$_\odot$~yr$^{-1}$
reproduces best their observations. However the total ejected mass predicted 
by \citet{Guerrero2018} is $8\times10^{-4}~$M$_\odot$.

All of our models predict that most of the mass in dust is present in the outer shell, which is located between the inner disrupted disk and the outer relic, H-rich nebula.
The reasons for this situation might be a combination of physical effects.
Radiation pressure over the grains will cause them to be accelerated pushing them
away from the CSPN and the complex hydrodynamic processes
occurring in the H-deficient clumps and filaments of A\,30 and A\,78 
\citep[see discussion section in][]{Fang2014} might drag the dust along
with the gas depending on the coupling degree. However,
one might also need to take into account the destruction of dust close to the
CSPN \citep[see][]{Borkowski1994}.
Within 1000~yr after the born-again event in A\,30 and A\,78 it is difficult to 
assess which effect has dominated over the other. This could be only explored 
by future radiation-hydrodynamic simulations that include dust physics and a 
detailed evolution of the stellar wind parameters and UV flux from stellar
evolution models tailored to VLTP events.

\subsection{The born-again {\it versus} the nova-like scenario}

The born-again scenario described in Section~1 has been challenged by the nebular abundance measurements of their H-deficient material. Using optical spectroscopic observations \citet{Wesson2003,Wesson2008} estimated that the C/O ratio by mass of the H-poor material in A\,30 and A\,58 have values between 0.06 and $\lesssim0.3$, whilst the born-again scenario calculations predict C/O$\gtrsim 1$ \citep[see table~2 in][]{MM2006}. Following this discrepancy, \citet{Lau2011} suggested that born-again PNe were better explained by nova-like eruptions, which predict C/O values around 0.05--0.60 \citep[see, e.g.,][]{Starrfield1998}, instead of the VLTP scenario.
However, modeling the optical and IR observations of A\,30 with {\sc cloudy} we have demonstrated that around 40 per cent of $M_\mathrm{VLTP}$ might have coagulated into C-rich dust, which is not accounted for in the nebular C/O estimations described above.

Using the nebular abundances by mass fraction of C and O reported for knot 1 in A\,30 by \citet{Wesson2003} \citep[see also table~1 in][]{Lau2011}, that is $X_\mathrm{C}$=0.07 and $X_\mathrm{O}$=0.27, we can estimate their contribution to the gas as 
\begin{equation}
M_\mathrm{gas,C}=X_\mathrm{C}\times M_\mathrm{gas} = 3.1\times10^{-4}~\mathrm{M}_{\odot}
\end{equation}
\noindent and
\begin{equation}
M_\mathrm{gas,O}=X_\mathrm{O}\times M_\mathrm{gas} = 1.2\times10^{-3}~\mathrm{M}_{\odot},
\end{equation}
\noindent respectively. On the other hand, the mass of carbon in the dust is exactly the dust mass determined by our best model, $M_\mathrm{dust}$. This follows from the fact that amorphous carbon and graphite are allotropic forms of carbon, that is, they are formed only of carbon. Finally, we can estimate the C/O mass ratio as
\begin{equation}
    \mathrm{C/O} = \frac{M_\mathrm{gas,C} + M_\mathrm{dust}}{M_\mathrm{gas,O}} \geq 1.27,
\end{equation}
the later adopting the lower limit to the dust mass.

The revised value of the C/O ratio is then consistent with the born-again scenario predictions instead of a nova-like event. The possible role of the binary companion of the CSPN of A\,30 in the ejection of the H-poor material is then downgraded, although it might still have helped shaping the axisymmetric ejecta.

\section{SUMMARY}

We presented the analysis of IR observations of the born-again PNe
A\,30 and A\,78. These two PNe have experienced an explosive event 
that ejected
processed material inside their old nebular structures. Previous
optical observations have shown that, in both cases, the
H-deficient material is distributed in a disk and jet close to
their CSPN plus a collection of clumps and filaments distributed
more or less homogeneously inside their old, hydrogen-rich PNe.

IR images and spectra covering the 3--160~$\mu$m wavelength range were
used in conjunction with the photoionization code {\sc cloudy} to study the
distribution of carbon-rich dust in A\,30 and A\,78 and to
characterize its properties. Our findings can be summarized as:

\begin{itemize}
  
\item We confirm previous suggestions that the dust in A\,30 and A\,78
  is not only distributed in a disk close to the star. Mid-IR images
  obtained from {\it Spitzer} IRAC and {\it WISE} show that dust is
  present up to distances $\sim40$~arcsec from their CSPN. Comparison
  with optical and X-ray images show that the dust is spatially
  correlated with the H-deficient clumps (including the jet
  features) and filaments inside the old H-rich PNe, and that it coexists
  with the recently-powered X-ray-emitting hot bubbles ($T_\mathrm{X}\lesssim$10$^{6}$~K).
  
\item Analysis of the near- and mid-IR images suggest that the dust
  present in the disks of A\,30 and A\,78 have different properties
  than their jet features. In particular, comparison between $H$ and
  $K$ band images of A\,30 with those obtained with {\it Spitzer} IRAC
  show that the jet starts to be detected at wavelengths longer than 
  3~$\mu$m.

\item The spectral maps obtained with the {\it Spitzer} IRS observations show
  the first imaging evidence of amorphous carbon in A\,78. 
  These are mapped through the 6.4 and 8.0~$\mu$m broad spectral features.
  Its spatial coincidence with the X-ray emission supports the charge exchange reactions at the surface of these carbon-rich structures as the mechanism for the production of super soft X-ray in born-again PNe. 
  We detect the N\,{\sc v} 29.43~\AA\, line in absorption in A\,30, A\,58 and A\,78, which has been suggested to originate in the $10^{5}$~K gas resulted from the mix of the X-ray-emitting gas and the ionized material.
  Thus, we predict that as well as the other two objects, A\,58 might have started to develop a hot bubble.

\item A variety of {\sc cloudy} models including a number of dust species
  were used to study their contribution in different regions of the IR SED. Models with single
  dust species are not able to reproduce simultaneously the SED in the mid- and far-IR.
  A combination of different carbon-rich species need to be
  taken into account, for example, our best model includes the contribution from fluffy 
  amorphous carbon and graphite. The advantage of using {\sc cloudy} is that we can 
  simultaneously fit the nebular and dust properties for a more complete description of
  born-again PNe and produce improved estimates of dust temperatures. 
  The dust temperatures predicted by our models are in the 
  40--230~K range, in stark contrast with previous estimates of 1000~K.

\item Our best model for the nebular and IR properties of A\,30 was obtained by adopting a disk structure near the CSPN with the contribution of an outer spherical region, gas and dust is included in both regions. The total estimated gas is $(4.41^{+0.55}_{-0.14})\times10^{-3}$~M$_\odot$ with a very similar total mass of dust of $(3.20^{+3.21}_{-2.06})\times10^{-3}$~M$_\odot$. 
The model requires a small fraction $<$5\% of the dust to be present in the disk
with only the contribution from small grains (0.001--0.03~$\mu$m).  
Most dust is in the outer shell, specifically the big grains (0.06--0.20~$\mu$m). 
This situation shows the complex interactions dragging the dust away from the CSPN which might involve radiation pressure, hydrodynamic processes and dust destruction.

\item If the total mass ejected in the VLTP of A\,30 is
$M_\mathrm{VLTP}=M_\mathrm{dust} + M_\mathrm{gas}=[7.61^{+3.76}_{-2.20}]\times10^{-3}$~M$_\odot$,
then we estimated a mass-loss rate of $\dot{M}_\mathrm{VLTP}\approx[5-60]\times10^{-5}$~M$_{\odot}$~yr$^{-1}$.
Due to the very likely porous (chain-like) structure of amorphous carbon
grains, we favor lower values for both $M_\mathrm{VLTP}$ and $\dot{M}_\mathrm{VLTP}$, 
which are otherwise consistent with previous estimates for the born-again PN HuBi\,1.

\item We have shown that the C mass in the H-poor ejecta of A\,30 is larger than the O mass once that the C atoms trapped into dust are taken into account. This result is consistent with the predictions of the born-again scenario and contradicts the expectations of a nova-like event suggested previously.

\end{itemize}

Born-again PNe are complex entities with only a few cases known 
which remain to be unveiled. Future SOFIA HAWC$+$ observations
at longer wavelengths could be used to
constrain the contribution from graphite in evolved born-again PNe.
Similar studies of other known born-again PNe will be pursued to assess the C/O mass ratio discrepancy pointed out in previous works.
Future radiation-hydrodynamic numerical simulations including dust-related physics, following in detail the evolution of the CSPN, might be able to shed light into the 
complex phenomena occurring in born-again PNe.

\section*{Acknowledgements}

The authors are thankful to the anonymous referee for a detailed report of our manuscript which improved its presentation and clarity.
The authors thank O.\,Gonz\'{a}lez-Mart\'{i}n (IRyA-UNAM) for reducing
the CanariCam data. 
The authors thank funding by
Direcci\'{o}n General de Asuntos del Personal Acad\'{e}mico (DGAPA) of the Universidad 
Nacional Aut\'{o}noma de M\'{e}xico (UNAM) project IA100720. 
JAT thanks Fundaci\'{o}n Marcos Moshinsky (Mexico).
VMAGG acknowledges support from the Programa de Becas 
posdoctorales funded by DGAPA UNAM. JBRG, SED and PJH thank Consejo
Nacional de Ciencias y Tecnolog\'{i}a (CONACyT) M\'{e}xico for research
student grants.
MAG acknowledges support of the Spanish Ministerio de Ciencia, Innovaci\'on 
y Universidades (MCIU) grant PGC2018-102184-B-I00.
GRL acknowledges support from CONACyT (grant 263373) and PRODEP (Mexico).
DAGH acknowledges support from the State Research Agency (AEI) of the 
Spanish Ministry of Science, Innovation and Universities (MCIU) and 
the European Regional Development Fund (FEDER) under grant 
AYA2017-88254-P.

Based on observations made with the instruments NOTCam at 
the Nordic Optical Telescope (NOT) and CanariCam at the 
Gran Telescopio Canarias (GTC), installed in the Spanish 
Observatorio del Roque de los Muchachos of the Instituto 
de Astrofísica de Canarias, in the island of La Palma. 
NOT is owned in collaboration by the University of Turku 
and Aarhus University, and operated jointly by Aarhus 
University, the University of Turku and the University 
of Oslo, representing Denmark, Finland and Norway, the 
University of Iceland and Stockholm University. This work makes use
of {\it Spitzer} IR observations which was operated by the 
Jet Propulsion Laboratory, California Institute of Technology 
under a contract with NASA. Support for this work was provided 
by NASA through an award issued by JPL/Caltech. This work is partially based 
on observations obtained with {\it XMM–Newton}, an ESA science mission with instruments 
and contributions directly funded by ESA Member States and NASA.
This work has made extensive use of NASA’s Astrophysics Data System.

\section*{DATA AVAILABILITY}

The data underlying this article will be shared on reasonable request
to the corresponding author.





\appendix

\section{The presence of the mixing layer on born-again planetary nebulae}
\label{sec:appA}

The temperature structure of wind-blown bubbles results in a somewhat 
complex distribution.
An X-ray-emitting hot bubble ($T_\mathrm{X}\gtrsim10^6$~K) 
is formed by the adiabatically-shocked stellar wind that pushes 
the photionized material ($10^{4}$~K). 
Numerical simulations have long predicted that, when in contact,
these two layers might experience a variety of physical processes. The 
photoionized material might experience hydrodynamical instabilities 
breaking into clumps and filaments \citep[e.g.,][and references therein]{Toala2011,Toala2018}
that are latter mixed with the hot bubble
producing gas at intermediate temperatures. This process can be enhanced if 
thermal conductivity is taken into account \citep{Soker1994,Weaver1977}.
As a result, a mixing (or conduction) 
layer with temperature $\sim10^{5}$~K between these two layers
will appear.

Conduction layers have been detected in other PNe by means of UV observations. 
Far Ultraviolet Spectroscopic Explorer (FUSE) observations of the 
O\,{\sc vi}~$\lambda\lambda$1032,1038 doublet have been reported in several PNe
\citep{Iping2002,Gruendl2004,Ruiz2013}. \citet{Fang2016} used {\it HST}
STIS observations to disclose the presence of the N\,{\sc v} 1239~\AA\, 
emission line in the Cat's Eye Nebula in comparison with the 
X-ray-emitting gas and the nebular layers, revealing for the first time
the distribution of the mixing layer in a PN.

\begin{figure}
\begin{center}
  \includegraphics[angle=0,width=\linewidth]{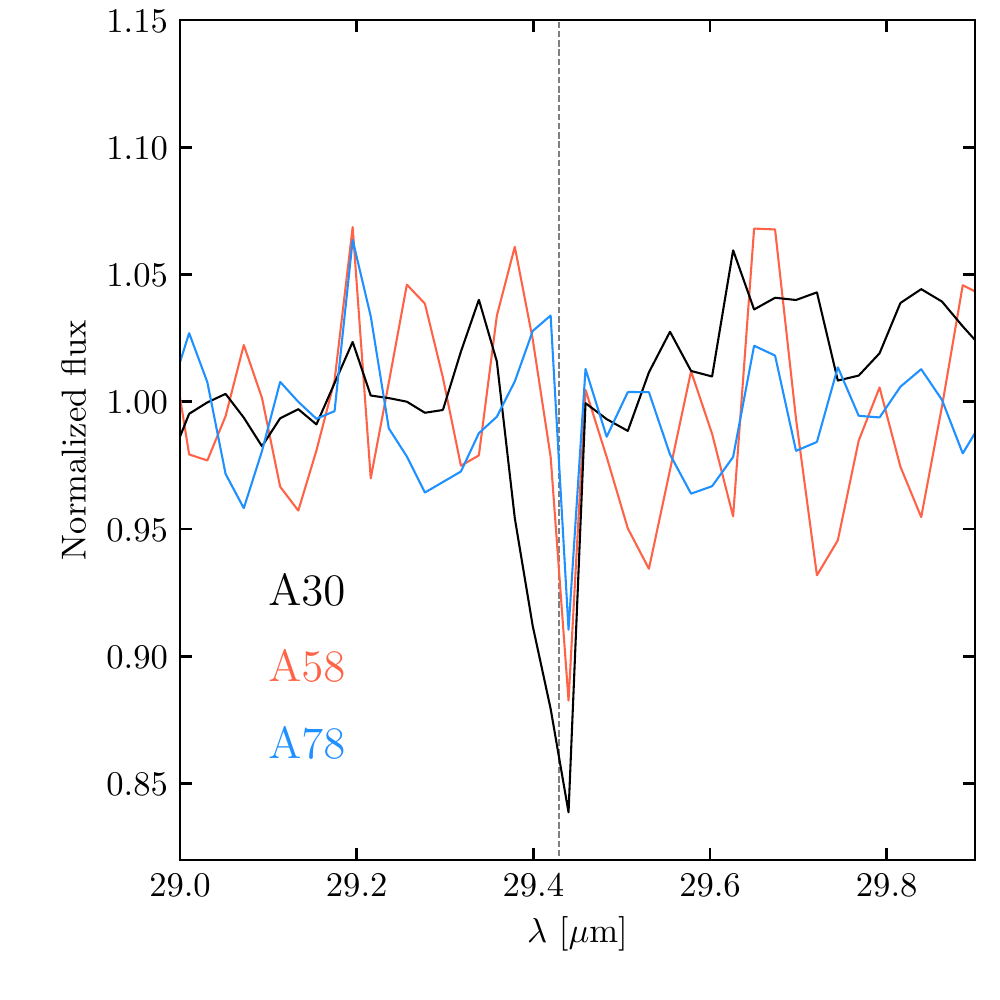}
\caption{Normalized flux of the high-resolution spectra of A\,30, A\,58 and A\,78
around the N\,{\sc v} absorption line at 29.43~$\mu$m (dashed line).}
\label{fig:N_V}
\end{center}
\end{figure}

The N\,{\sc v} 29.43~$\mu$m line is detected in absorption in the high-resolution
{\it Spitzer} IRS observations of the three born-again PNe presented here (see Fig.~\ref{fig:N_V}). We used {\sc iraf}
to estimate that the signal-to-noise (S/N) of the N\,{\sc v} line is 54, 33 and 38 
for A\,30, A\,58 and A78 with FWHM of 0.06, 0.03 and 0.02~$\mu$m, respectively. These where
then compared to the FWHM of the unambiguously detected O\,{\sc iv} at 25.88~$\mu$m which 
resulted in 0.05~$\mu$m (A\,30), 0.026~$\mu$m (A\,58) and 0.045~$\mu$m (A\,78). 
Thus, we can conclude that 
the N\,{\sc v} absorption line is definitely detected in A\,30 and A\,58, but questionably in A\,78.
However, the reported extended X-ray emission in A\,78 corroborates the detection of the
mixing layer through the N\,{\sc v} absorption line.

These results suggest that A\,58 might also harbor a hot bubble formed by the interaction of
its current fast wind with the dense structure around its CSPN. Future X-ray observations will
help assessing this suggestion.

\end{document}